\documentclass[aps,prl,twocolumns,showpacs]{revtex4}

\usepackage{amsmath,epsfig,amsfonts,epsfig,graphicx,
}

\begin{document}

\title{Localized Breathing Modes in Granular Crystals with Defects}
\author{G. Theocharis$^{1}$, M. Kavousanakis$^{2}$, P.~G. Kevrekidis$^{1}$, Chiara Daraio$^{3}$, Mason A. Porter$^{4}$, I.~G. Kevrekidis$^{2}$}
%\\
\affiliation{ 
$^1$ Department of Mathematics and Statistics, University of Massachusetts, Amherst MA 01003-4515, USA \\
$^2$ Department of Chemical Engineering, Princeton University, Princeton, NJ, 08544, USA \\
$^3$ Graduate Aerospace Laboratories (GALCIT), California Institute of Technology, Pasadena, CA 91125, USA \\
$^4$ Oxford Centre for Industrial and Applied Mathematics, Mathematical Institute, University of Oxford, OX1 3LB, UK
}

\begin{abstract}
We investigate nonlinear localized modes at light-mass impurities in a one-dimensional, strongly-compressed chain of beads under Hertzian contacts.  Focusing on the case of one or two such ``defects,'' we analyze the problem's linear limit to identify the system eigenfrequencies and the linear defect modes.  We then examine the bifurcation of nonlinear defect modes from their linear counterparts and study their linear stability in detail.  We identify intriguing differences between the case of impurities in contact and ones that are not in contact.  We find that the former bears similarities to the single defect case, whereas the latter features symmetry-breaking bifurcations with interesting static and dynamic implications. 
\end{abstract}


\maketitle

\section{Introduction}

In the present work, we investigate nonlinear localized modes resulting from configuration heterogeneity in granular crystals.  This entails a confluence of three key research themes: intrinsic localization through nonlinearity, wave propagation in granular chains, and localization through extrinsic disorder.

Intrinsic localized modes (ILMs), otherwise known as discrete breathers, have been a central theme for numerous theoretical and experimental studies over the past two decades \cite{camp04,flawil,macrev,Flach2007,aubrev}.  The original theoretical
proposal of ILMs in prototypical settings such as anharmonic nonlinear lattices
\cite{sietak,pa90} and the rigorous proof of their existence under fairly general conditions \cite{macaub} motivated numerous studies of such modes in a diverse host of applications, including
optical waveguides and photorefractive crystals \cite{photon}, the denaturation of the DNA double strand \cite{peyrard}, micromechanical cantilever arrays \cite{sievers}, nanomechanical resonators \cite{kenig09}, superconducting Josephson junctions \cite{orlando}, as well as Bose-Einstein condensates \cite{morsch1}, and electrical lattices \cite{lars1}, among many others.

One-dimensional (1D)  granular crystals, consisting of closely-packed chains of elastically interacting particles, have drawn considerable attention during the past few years.  This broad interest has arisen from the wealth of available material types/sizes and the ability to tune their dynamic response to encompass linear, weakly nonlinear, and strongly nonlinear regimes \cite{nesterenko1,sen08,nesterenko2,coste97}.  Such flexibility makes them perfect candidates
for many engineering applications, including shock and energy absorbing 
layers \cite{dar06,hong05,fernando,doney06}, actuating devices \cite{dev08}, and sound scramblers \cite{dar05,dar05b}.  Because of these possibilities, it is crucial to investigate the effects of defects (imhomogeneities, beads with different masses, etc.), allowing the observation of interesting physical responses such as fragmentation, anomalous reflections, and energy trapping \cite{dar06,hong05,fernando,doney06,Hascoet,hinch99,hong02,sen98,manciu99,dimershort,dimerlong}. 

It is well-known from solid-state physics that localized vibrations in linear lattices can arise from extrinsic disorder that breaks the discrete translational invariance of a perfect crystal lattice \cite{Maradudin,Lif}.
Such ``disorder'' is also well known to introduce interesting wave-scattering effects  \cite{marad}. This phenomenology arises in a wide host of physical applications, including superconductors \cite{andreev}, 
electron-phonon interactions \cite{tsironis}, light propagation in dielectric super-lattices with embedded
defect layers \cite{soukoulis}, defect modes in photonic crystals \cite{Joann}, and optical waveguide arrays~\cite{Pes_et99,Mor_et02,kipp}.  

In the present work, we aim to investigate the confluence of the preceding research themes by examining the interplay of ``disorder'', which induces localized modes, and nonlinearity in granular crystals. 
We notice in passing that the interaction of impurities with 
solitary waves or a continuous oscillatory signal in non-loaded (i.e., without precompression) monomer 
chains has been investigated numerically \cite{Hascoet,sen98} as well as 
experimentally in the recent work of \cite{Job}. In these studies, 
localized oscillations result from the presence of an impurity of lighter mass than the remaining chain particles, during the interaction of the impurity with either a solitary wave or a continuous oscillatory signal.
However, these localized oscillations were all transient, fading away as soon as the wave left the vicinity of the impurity. 

Here, by contrast, we examine \textit{long-lived} localized breathing oscillations, which form robust nonlinear localized modes (NLMs) induced by the impurities, 
in strongly-compressed granular chains.  We demonstrate that their frequency depends not only on experimental parameters such as the precompressive force and the constitution (material and size) of the impurity bead but also on the inherent nonlinearity of the system (i.e., the amplitude of the oscillations).   
We provide a detailed bifurcation and dynamical
analysis of a monoatomic chain with a single lower-mass bead (``impurity") and extend our considerations to monoatomic chains with a pair of lower-mass impurities.  We show that the wave dynamics in the case of nearest neighbor ``impurities'', which contains consecutive lower-mass beads without other intervening particles, differs substantially from that in the case of larger separations between impurities.  We focus on the case of next-nearest-neighbor impurities, revealing its rich bifurcation structure in which strongly asymmetric branches of solutions emerge through symmetry-breaking.  We monitor the dynamical manifestation of this bifurcation (and associated instability) and examine how it is affected
by a potential initial asymmetry in the impurity masses.

The remainder of our presentation is structured as follows.  We start by discussing the general theoretical setup of the homogeneous (no impurity) model.  We then perform linear and nonlinear analyses, first for monoatomic chains with a single impurity and then for monoatomic chains with a pair of impurities.  Finally, we summarize our findings and present some possible future directions.

\section{Monoatomic Granular Chain}

The interaction between two adjacent elastic spheres is well-known to be described by Hertz's law \cite{Landau}.  The relation between the force $F_0$ exerted on two identical spheres and the distance $\delta_{0}$ between their centers  results from geometric effects and is given by the nonlinear relation
\begin{equation}
	F_{0} = A\delta_{0}^{3/2}\,, \label{Hertz}
\end{equation}
where
\begin{equation}
	A = \frac{E\sqrt{2R}}{3\left(1-\nu^2\right)}\,,  \label{material}
\end{equation}
$R$ is the radius of the beads, $E$ is the material's elastic (Young's) modulus, and 
$\nu$ is the Poisson ratio of the bead material. 

The dynamics of a 1D chain composed of beads of a single type
(i.e., a monoatomic chain) is thus described by the following system of coupled
nonlinear ordinary differential equations \cite{nesterenko1}:
\begin{equation}
	M \ddot{u}_i = A[\delta_{0}+u_{i-1} - u_{i}]_+^{3/2} - A[\delta_{0}+u_{i} - u_{i+1}]_+^{3/2} \,,
%	A &= \frac{2E\left(\frac{R}{2}\right)^{1/2}}{3\left(1-\nu^2\right)}\,,	
\label{model}
\end{equation}
where $u_i$ is the displacement of the $i$th bead from its equilibrium position in the initially-compressed chain, $i \in \{2,\cdots,N-1\}$, and 
$M$ is the mass of the beads. The bracket $[s]_+$ of Eq.~(\ref{model}) takes the value $s$ if $s > 0$ and the value $0$ if $s \leq 0$, which signifies that adjacent beads are not in contact. 
 
In contrast with Ref.~\cite{Job}, which considered unloaded chains, we investigate strongly precompressed chains, in which $F_{0}$ takes large values.  Considering small amplitude displacements  
in comparison with the initial ones caused by the precompression force, namely
\begin{equation}
	\frac{|u_{i-1}-u_{i}|}{\delta_{0}}\ll1\, \label{approx}
\end{equation}
one can Taylor expand the forces in a power series, in which case 
keeping the  displacement terms to fourth order leads to the 
approximate (``$K_2-K_3-K_4$") model form
\begin{align}
	M\ddot{u}_i &= K_{2}(u_{i-1}-2u_{i}+u_{i-1}) \notag \\
	&+K_{3}\left((u_{i+1}-u_{i})^2-(u_{i-1}-u_{i})^2 \right)\notag \\
	&+K_{4}\left((u_{i+1}-u_{i})^3+(u_{i-1}-u_{i})^3 \right)\,, \label{K2K3K4}
\end{align}
where
\begin{equation}
	K_2 = \frac{3}{2}A\delta_{0}^{1/2}\,, \quad K_3=-\frac{3}{8}A\delta_{0}^{-1/2}\,, \quad K_4=\frac{3}{48}A\delta_{0}^{-3/2}\,. \label{K2K3K4coef}
\end{equation}
The equations of motion (\ref{K2K3K4}) are an example of the celebrated Fermi-Pasta-Ulam (FPU) model \cite{focus,fpupop,huang}.  If we restrict our consideration to very small amplitudes and velocities,
we can neglect all of the nonlinear terms from the equations of motion and keep only the harmonic ($K_2$) term.  The spectral band of the ensuing linear chain has an upper cutoff frequency of $\omega_{m}=\sqrt{4K_2/m}$, which corresponds to the lattice vibration in which the neighboring particles oscillate out of phase.

One of the remarkable features of gradually introducing nonlinearity is that its interplay with discreteness leads to the emergence of localized modes even in the absence of any inhomogeneity.  Such ILM solutions are generic features of a large class
of Hamiltonian 
lattices (which includes the FPU model) \cite{Flach2007}.  ILMs have been studied 
extensively in monoatomic FPU chains \cite{DBinFPU_F}.

One of the canonical mechanisms for the generation of these nonlinear 
localized modes, is the 
modulational instability (MI) of the band edge plane wave. A detailed analysis 
of this instability (bifurcation)
has shown that the MI of the upper cutoff mode manifests itself 
when the following inequality holds (see Sec. $4.3$ of \cite{Flach2007} and references therein)
\begin{equation}
3K_2K_4-4K_{3}^2>0.
\label{ineq}
\end{equation}
Using Eqs. (\ref{K2K3K4coef}) one can show that the above mentioned inequality 
doesn't hold in our setting, which, in turn, indicates that small 
amplitude ILMs bifurcating from the upper band mode don't exist in 
monoatomic granular crystals.
(The existence of dark discrete breathers or large amplitude DBs remains an 
interesting open question for future investigations).

\section{Monoatomic Granular Chain with an Impurity}

\subsection{Model}

Consider a 1D monoatomic chain of beads that contains an impurity at the $k$th site.  Suppose that the impurity (the $k$th bead) is made of the same material as the other particles but has a different radius. 
In particular, we choose the homogeneous host chain composed of spherical stainless steel beads of non-magnetic, 316 type (which have elastic modulus $E = 193$ GPa, Poisson ratio $\nu = 0.3$, and density $\rho = 8027.17$ kg/m$^3$ \cite{316}) of radius $R = 4.76$ mm and a spherical steel impurity bead with some other radius.  We treat the radius of the impurity as a free parameter, though in most cases we will use the radius $r = 0.8R$. We also suppose that the granular chain is compressed by an experimentally-accessible static force of $F_{0} = 22.25$ N.

The presence of the impurity bead in the chain gives rise to a mass defect and leads to changes in the force constants that destroy the discrete translational symmetry of the crystal.  Recalling that the precompressive static force $F_{0}$ induces an initial displacement $\delta_{0}$ 
between neighboring spheres of the same diameter and defining $\delta_{1}$ to be the displacement that it induces between neighboring spheres of different diameter, equations of motion (\ref{model}) at sites $k-1$, $k$, and $k+1$ become
\begin{align}
	  M\ddot{u}_{k-1} &= A_1[\delta_{0}+u_{k-2} - u_{k-1}]_+^{3/2} - A_2[\delta_{1}+u_{k-1} - u_{k}]_+^{3/2} \,, \notag \\
  	m\ddot{u}_{k} &= A_2[\delta_{1}+u_{k-1} - u_{k}]_+^{3/2} - A_2[\delta_{1}+u_{k} - u_{k+1}]_+^{3/2} \,, \notag \\
 	 M\ddot{u}_{k+1} &= A_2[\delta_{1}+u_{k} - u_{k+1}]_+^{3/2} - A_1[\delta_{0}+u_{k+1} - u_{k+2}]_+^{3/2} \,, \notag \\
  	A_{1} &= \frac{2E\left(\frac{R}{2}\right)^{1/2}}{3\left(1-\nu^2\right)}\,,	\qquad
	A_{2} = \frac{2E\left(\frac{R r}{R + r}\right)^{1/2}}{3\left(1-\nu^2\right)}\,,	
	\label{model1}
\end{align}
where $m$ and $r$ are the mass and the radius of the impurity bead while $M$ 
and $R$ denote the mass and the radius of the remaining beads.

\subsection{Harmonic Potential Approximation (Linear Analysis)}

To understand the underlying linear spectrum of the problem, we linearize (\ref{model1}) about the equilibrium state in the presence of precompression.  This yields
\begin{align}
  	m\ddot{u}_{i} &= C_1(u_{i-1} - 2u_{i} - u_{i+1}) \,, \quad  i\notin \{k-1,k,k+1\} \notag \,,  \\
  	M\ddot{u}_{k-1} &= C_1(u_{k-2} - u_{k-1}) - C_2(u_{k-1} - u_{k}) \,, \notag \\
  	m\ddot{u}_{k} &= C_2(u_{k-1} - 2u_{k} - u_{k+1}) \,, \notag \\
  	M\ddot{u}_{k+1} &= C_2(u_{k} - u_{k+1}) - C_1(u_{k+1} - u_{k+2}) \,, \notag \\
  	C_{1} &= \frac{3}{2}A_1\delta_{0}^{1/2}\,,	\qquad
	C_{2} = \frac{3}{2}A_2\delta_{1}^{1/2}\,.	
	\label{model2}
\end{align}

Seeking stationary solutions with frequency $\omega$, we substitute
\begin{equation}
	u_{j} = \upsilon_{j}e^{i\omega t}  \label{ansatz}
\end{equation}
for all $j$ into Eqs.~(\ref{model2}) and obtain the following eigenvalue problem:
\begin{equation} \label{eval}
	\omega^{2}M 
\left(
\begin{array}{ccccccc}
1 & 0 & &\ldots &  & 0&0\\
0 & 1 & &\ldots &  & 0&0\\
\vdots  &\vdots & \ddots & & & \vdots& \vdots  \\
0 & 0& \ldots & \frac{m}{M}& \ldots & 0&0\\  
\vdots  &\vdots & & &\ddots  & \vdots & \vdots\\
0 & 0 & &\ldots & & 1& 0 \\
0 & 0 & &\ldots & & 0& 1
\end{array} \right)\left(
\begin{array}{cccccc}
\upsilon_1 & \\
\upsilon_2 & \\
\vdots & \\ 
\upsilon_k & \\
\vdots & \\
\upsilon_{N-1} & \\
\upsilon_N &
\end{array} \right)
=C_1 \mathbf{V}\left(
\begin{array}{cccccc}
\upsilon_1 & \\
\upsilon_2 & \\
\vdots & \\ 
\upsilon_k & \\
\vdots & \\ 
\upsilon_{N-1} & \\
\upsilon_N &
	\end{array} \right)\,,
\end{equation}
where 
\begin{equation*}
	\mathbf{V} =
\left(
\begin{array}{ccccccc}
1 & -1 & 0 & & \ldots & 0& 0\\
-1 & 2 & -1 & & \ldots & 0& 0\\
\vdots & \vdots & \ddots & && \vdots &\vdots \\
0 & 0&  \ldots & \mathbf{C}& \ldots & 0 &0\\
\vdots  &\vdots & & &\ddots  & \vdots & \vdots \\
0 & 0 & &\ldots & -1& 2& -1 \\
0 & 0 & &\ldots & &-1& 1\\
	\end{array} \right)\,,
\end{equation*}
\begin{equation*}
	\mathbf{C} = 
	\left(
\begin{array}{ccccc}
-1 & 1+\frac{C_2}{C_1} & -\frac{C_2}{C_1} & 0& 0\\
0 & -\frac{C_2}{C_1} & 2\frac{C_2}{C_1} & -\frac{C_2}{C_1}& 0\\
0 & 0&  -\frac{C_2}{C_1}& 1+\frac{C_2}{C_1}& -1
	\end{array} \right)\,.
\end{equation*}

\begin{figure}[tbp]
\includegraphics[width=8cm]{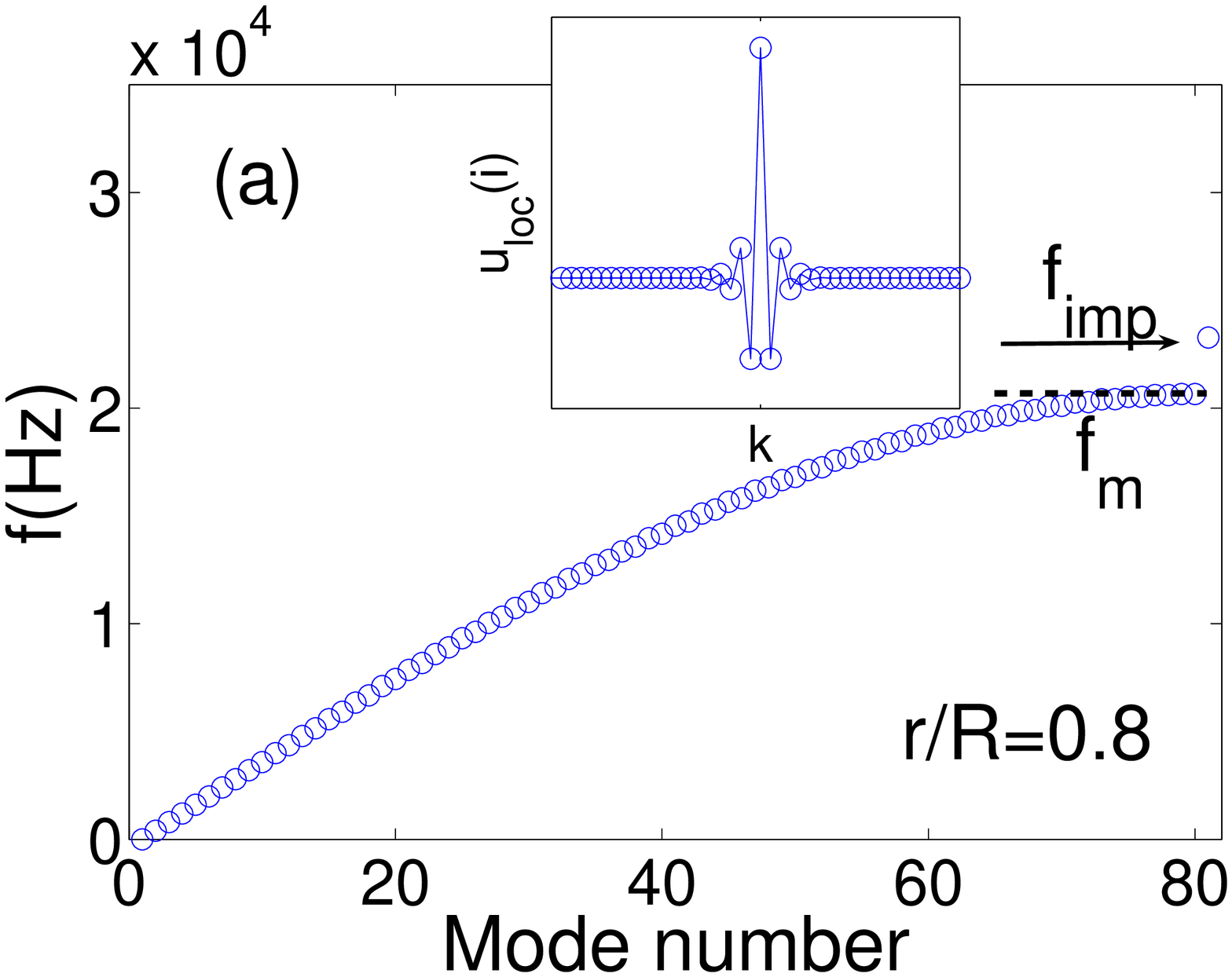}
\includegraphics[width=8cm]{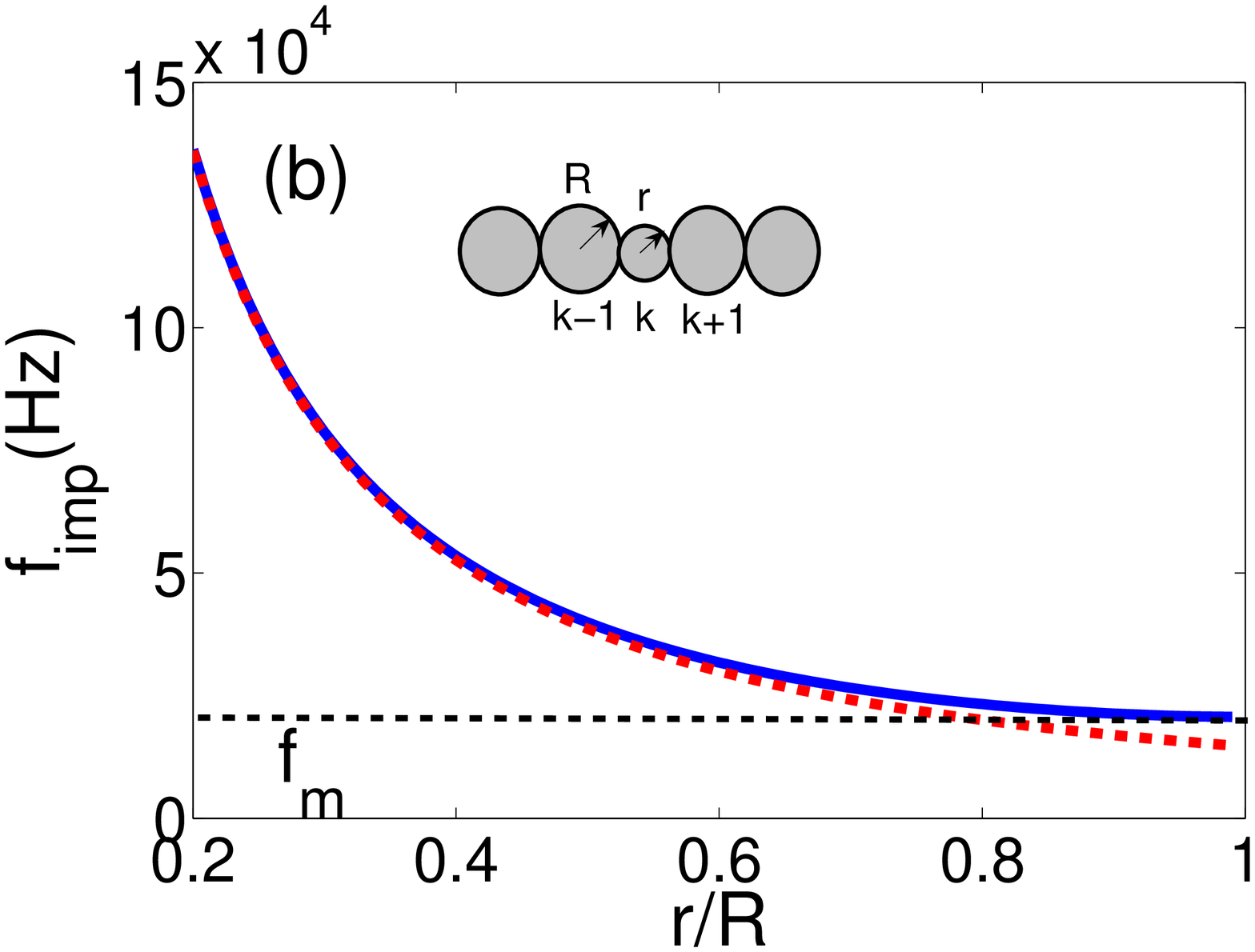}
\caption{(Color online) (a) The normal mode frequencies of the crystal in the presence of the single lighter-mass defect.  The presence of the impurity bead leads to the manifestation of a localized mode (see inset) with frequency above the band of the perfect (monoatomic) host crystal. (b) Numerically-obtained frequency (solid curve) of the localized mode ($f_{imp}$) as a function of the radius ratio $r/R$ compared with the analytical prediction (dashed curve).  As $r/R\rightarrow1$, one sees that $f_{imp}\rightarrow f_{m} \approx 20.67$ kHz (the horizontal dashed line) and the deviation between the two curves becomes larger. 
}
\label{1defectlinear}
\end{figure}

The eigenvalue problem (\ref{eval}) determines the spectrum of the extended phonon excitations and of the localized defect mode centered at the impurity site.  In general, the presence of impurity beads can create two types of vibrational modes: 
\begin{enumerate}
	\item{\textit{Resonance modes}, when the mass of the impurity bead is larger than the mass of the rest ($m>M$).}
	\item{\textit{Localized modes}, in the opposite case ($m<M$).}
\end{enumerate}	 
Each resonance mode has a frequency within the range of frequencies that constitute the phonon band of the homogeneous host crystal and has a vibration amplitude that is larger in the vicinity of the impurity bead.  Each localized mode, on the other hand, has a frequency $f_{imp}$ that lies above the band of the normal modes frequencies of the homogeneous host crystal and, as shown in Fig.~(\ref{1defectlinear}), has a vibration amplitude that is large at the impurity site but decreases very rapidly with increasing distance.  Reference \cite{Job} used multiple-scale analysis to obtain the analytical approximation
\begin{equation}
	f_{a}\approx \frac{\sqrt{3}}{2\pi}\frac{A_2^{1/3}F_{0}^{1/6}}{m^{1/2}}
\label{Job_prediction}
\end{equation}
for $f_{imp}$.  It is important to observe that our setting is very different from the one in \cite{Job}.  In particular, the experimental setup in \cite{Job} has no precompression, so a travelling envelope moves over the defect site.  As discussed in \cite{Job}, this pulse acts as a ``local precompression force'' as it travels, which makes the system weakly nonlinear locally and results in localized oscillations at the impurity site.  By contrast, in the present setting, the chain is strongly compressed by a static force, which acts globally in a
nonlinear fashion and allows localized modes to be maintained indefinitely without further external excitations after they are excited initially (by, e.g., an actuator or the impact of a striker particle).
The above distinction between local and global forces is the key difference that leads to very long-lived localized oscillations around the impurity bead.
Such oscillations can last arbitrarily long in principle, but in laboratory experiments the presence of dissipative effects will eventually result in the attenuation of these localized modes \cite{CarreteroPRL}.

As illustrated in Fig.~\ref{1defectlinear}(b), we find excellent agreement between the analytical expression of Eq.~(\ref{Job_prediction}) and the frequency of the localized mode obtained by the eigenvalue system (\ref{eval}) up to radii ratios $\frac{r}{R} = 0.6$ (see also Fig.~5 of \cite{Job}).  In particular, for the material parameters and the precompressive force discussed above, an impurity bead of radius $r = 0.6R$ (for which $\frac{m}{M} \approx 0.216$) yields a localized mode with frequency $f_{imp} \approx 31.76$ kHz, whereas Eq.~(\ref{Job_prediction}) predicts $f_{a} \approx 30$ kHz.  On the other hand,
for $r = 0.8R$ (implying that $\frac{m}{M} \approx 0.512$), the eigenvalue system (\ref{eval}) gives $f_{imp} \approx 23.28$ kHz and Eq.~(\ref{Job_prediction}) predicts $f_{a} \approx 20.03$ kHz.  To provide additional context, we remark that the upper cutoff frequency of the precompressed homogeneous host crystal is 
given by $f_{m}=\frac{1}{2\pi}\sqrt{\frac{4K_2}{m}} \approx 20.67$ kHz. It is clear that
the analytical expression in Eq.~(\ref{Job_prediction}) is expected to be a good approximation only for $m \ll M$. Otherwise, one has to use the numerically-obtained frequency $f_{imp}$.  Additionally, as the radius of the impurity bead becomes smaller, the difference between the frequency $f_{imp}$ of the impurity-induced localized mode and the upper cutoff frequency becomes larger.
Put another way $f_{imp}\rightarrow f_{m}$ as $r/R\rightarrow1$ (as shown in Fig.~\ref{1defectlinear}(b)), while the localized mode becomes concomitantly more extended.

\subsection{Continuation and Stability Analysis}

In the previous section, we examined the linear response of the granular crystal in the presence of a single defect.  We now study Eq.~(\ref{model1}) directly, to examine the nonlinear behavior of the chain.  From a physical perspective, 
this will allow us to examine the interplay of nonlinearity and ``disorder''.  When a perfect nonlinear lattice supports ILMs, the presence of impurities can drastically change the properties of such localized modes, resulting in 
interesting phenomena such as the presence of asymmetric impurity modes in 
nonlinear lattices with a light-mass defect \cite{Kiv1}, or the existence of stable, nonlinear, heavy-mass impurity modes \cite{Kiv2}.  As we have mentioned, our nonlinear lattice does not support ILMs 
[see the inequality (\ref{ineq})], so the aforementioned phenomena are not expected to be present. However, the linear localization at the defect in conjunction with nonlinearity can result in the presence of robust NLMs.

More specifically, it is important to consider whether the nonlinearity of the chain can support 
the existence of localized modes with frequencies $f_{m} < f < f_{imp}$. In the linear limit, the chain does not support vibrations with frequencies in this regime.  To answer that question, we perform 
a parameter continuation starting from the linear localized mode and systematically changing (in small steps) the frequency from $f_{imp}$ towards the upper cutoff $f_{m}$. For each of the intermediate frequencies, we identify NLMs
to high precision, via a Newton method in phase space, using free boundary conditions and chains with $N=79$ beads.  
In order to identify the relevant branch of solutions, we use as an initial guess the localized impurity-induced mode (see insert of Fig.~\ref{1defectlinear}(a)), as this was obtained from the linear eigenvalue problem (\ref{eval}). 
The momenta of all the sites can be fixed to zero, 
following \cite{marin_aubry}, due to the 
time reversibility of the system. 
For details of this continuation method, 
see Ref.~\cite{Flach2007} and references therein. 

\begin{figure}[tbp]
%\centering 
%\includegraphics[width=8cm]{bifurcation_r_8_10.eps}
%\includegraphics[width=6cm]{contin1defect_b_22899_1.eps}
%\includegraphics[width=6cm]{maxeigenvalue_8_10.eps}
\includegraphics[width=8cm]{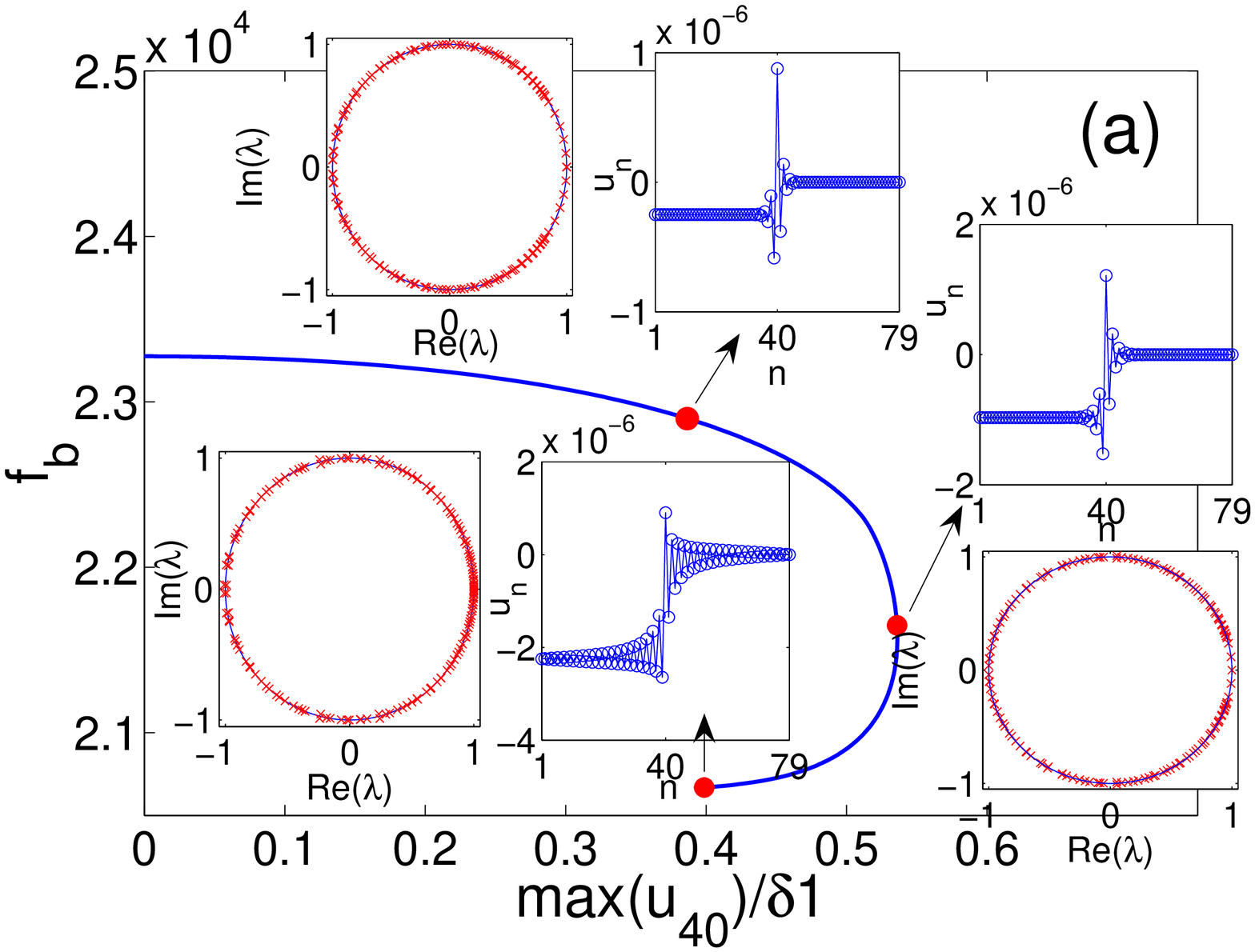}
\includegraphics[width=6cm]{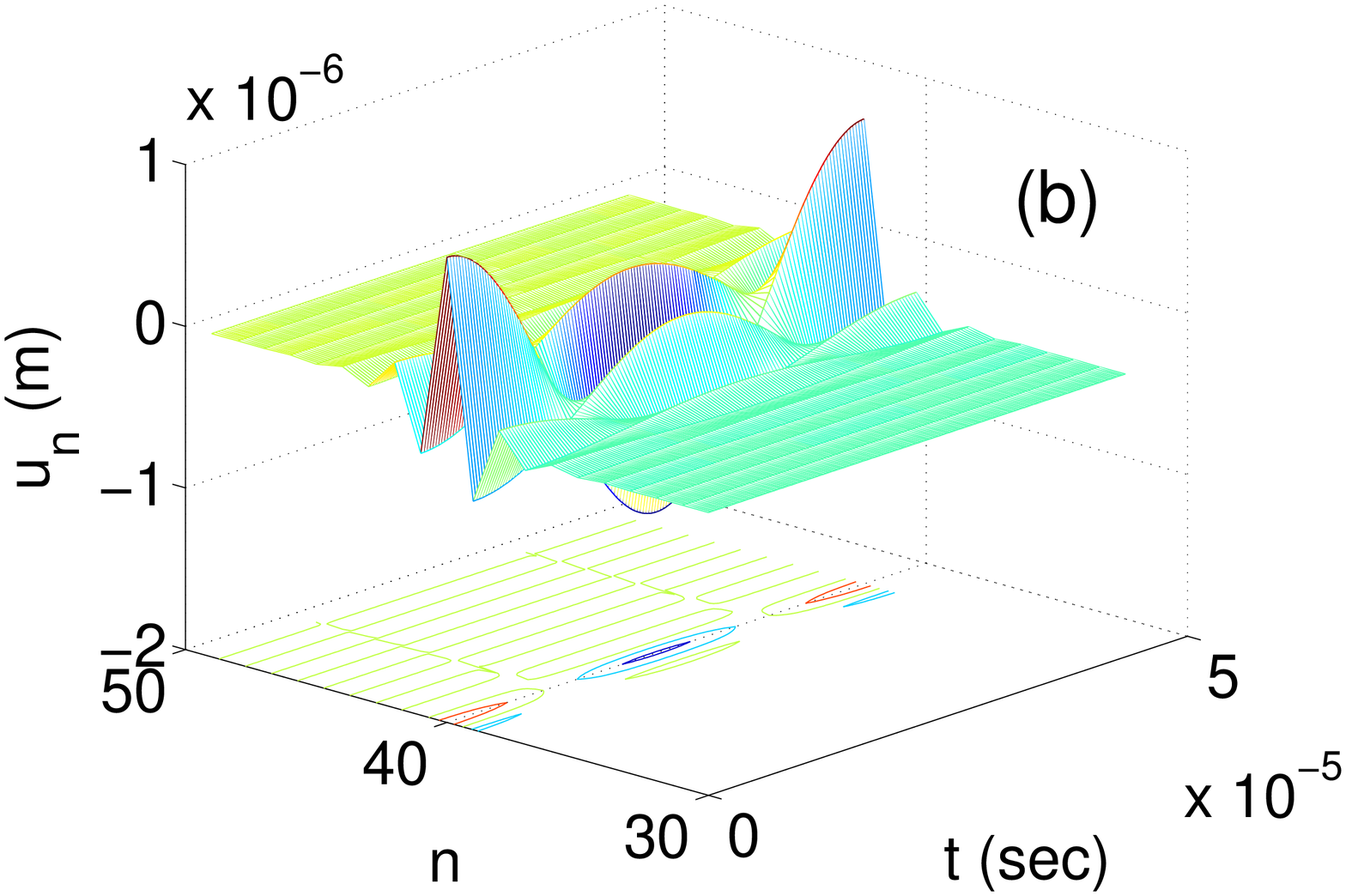}
\includegraphics[width=6cm]{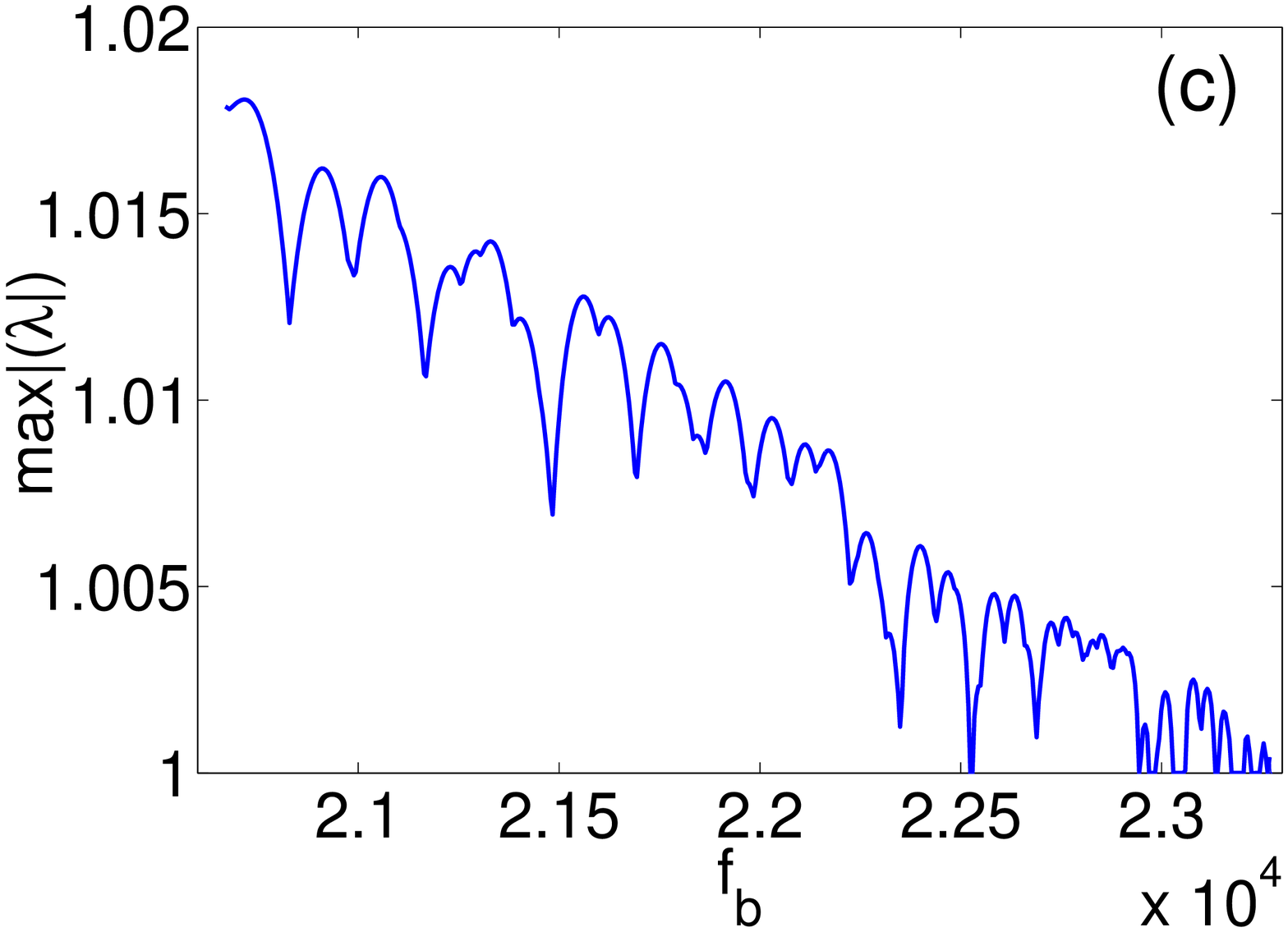}
\caption{(Color online)(a) Continuation diagram of the nonlinear impurity modes.  Insets: The spatial profiles and the corresponding locations of the Floquet multipliers $\lambda$ in the complex plane of the localized modes with frequencies $f_1 = 22.9$ kHz, $f_2=21.65$kHz, and $f_m = 20.67$ kHz. (b) One period of the spatiotemporal evolution of the localized mode with frequency $f_{1} = 22.9$ kHz.  (c) The maximum of the absolute value of Floquet multipliers as a function of the frequency $f_{b}$ of the nonlinear impurity mode.
}
\label{contin}
\end{figure}

We show the results of the continuation in Fig.~\ref{contin}, which allowed us to obtain localized solutions for all frequencies $f \in [f_{m}, f_{imp})$.  In insets of Fig.~\ref{contin}(a), we show three examples of these solutions; they have frequencies $f_1 = 22.9$ kHz, $f_2 = 21.65$ kHz, and $f_m = 20.67$ kHz.  We examined the stability of these localized modes by computing their Floquet multipliers $\lambda_{j}$, which describe the behavior of trajectories near the periodic solution.  We show the locations in the complex plane of the Floquet multipliers
for the three NLM profiles in insets of Fig.~\ref{contin}(a).  As is well-known, if all eigenvalues $\lambda_{j}$ have unit magnitude, then the localized periodic solution is linearly stable.  However, if $|\lambda_{j}|>1$ for some $j$, then a perturbation along the corresponding eigenvector $\mathbf{e_{j}}$ grows by the factor $|\lambda_{j}|$ after one complete period.  In Fig.~\ref{contin}(b), we show one period of the spatiotemporal evolution of the localized mode with frequency $f_{1}$.  In Fig.~\ref{contin}(c), we show the absolute value of the maximal eigenvalue, which is associated with the
instability growth rate.  

The family of the localized solutions was found to exhibit an oscillatory instability. In general, oscillatory instabilities may
arise either due to collision of Floquet multipliers associated with two extended eigenvectors, or between ones associated with an extended and one localized. 
In our case, a careful study of the unstable Floquet multipliers and the corresponding eigenvectors reveals
that the oscillatory instabilities are caused by the collision of extended modes belonging to the two arcs of overlapping continuous phonon spectrum of the Floquet
matrix~\cite{Finitesize}. During the continuation of the solutions, typically $3-5$ quadruplets of eigenvalues abandon the unit circle after the collision but return to it jointly soon afterwards in parameter space. As discussed in Ref.~\cite{Finitesize}, the strength of 
this kind of instabilities should depend on the system size (i.e., the number of beads in the chain) and vanishes in the limit of an infinite system. The deviations of the unstable eigenvalues from the unit circle are only up to $0.02$, and numerical integration of the nonlinear impurity modes up to times $100 T$ (where $T$ is their period) reveals their robustness.

It is relevant to also note that, among all the Floquet multipliers, 
two pairs are always located at $+1$ in the complex plane. 
One of them, the  so-called phase mode, describes a rotation of the overall 
phase of the breather, while the second one is due to the conservation of the 
total mechanical momemtum, an additional integral of motion of the FPU chains.
As one can see in the insets of Fig.~\ref{contin}(a), the spatial profiles of the NLMs have interesting structure.  In particular, they are characterized by a kink-shaped distortion of the chain, which is caused by the asymmetry in the interaction potential (see Sec. $4.1.4$ of \cite{Flach2007}). 
This asymmetry is evident in the $K_2-K_3-K_4$ approximation of the model and it arises directly from the fact that $K_3\neq 0$.  
The NLM can be thus viewed as a localized vibration which induces this kind of distortion into the granular crystal. As one approaches the edge of the phonon band 
(i.e., as $f_{b}\rightarrow f_{m}$), the NLMs become more extended and gradually approach their extended (plane wave) linear counterparts at the upper band of the 
linear spectrum.  In this limit, the dc distortion and the maximum of the absolute value of the associated Floquet mulipliers also increase.

\subsection{Excitation of Nonlinear Impurity Modes}

In the previous section we demonstrated that nonlinear localized modes exist in the gap between the band of phonon modes of the perfect crystal and the localized impurity mode.  This demonstrates that their frequency depends not only on the precompressive force and the material parameters of the defect chain \cite{Job} (as is the case in the linear limit), but also on the amplitude displacement of the impurity bead---in other words, on the strength of the nonlinearity.

In this light, we showcase in the present section what we believe is the easiest way 
to observe these modes.  Our method is based on the use of a simple localized initial 
excitation.  At time $t = 0$, we displace the bead at impurity site $k$ by an amount that is strong enough to ensure that the nonlinear terms are no longer negligible in the 
corresponding equations of motion.  Meanwhile, we keep all of the remaining sites at rest.  We then integrate the equations of motion (\ref{model1}) using a fourth-order 
Runge-Kutta numerical scheme.  We expect part of the initially-localized energy excitation to spread among the other sites.  In order to avoid back-scatter of emitted waves, 
we consider a chain with $N = 500$ beads.

\begin{figure}[tbp]
%\centering 
%\includegraphics[width=8cm]{generationa.eps}
%\includegraphics[width=8cm]{generation_b_d1.eps}
\includegraphics[width=8cm]{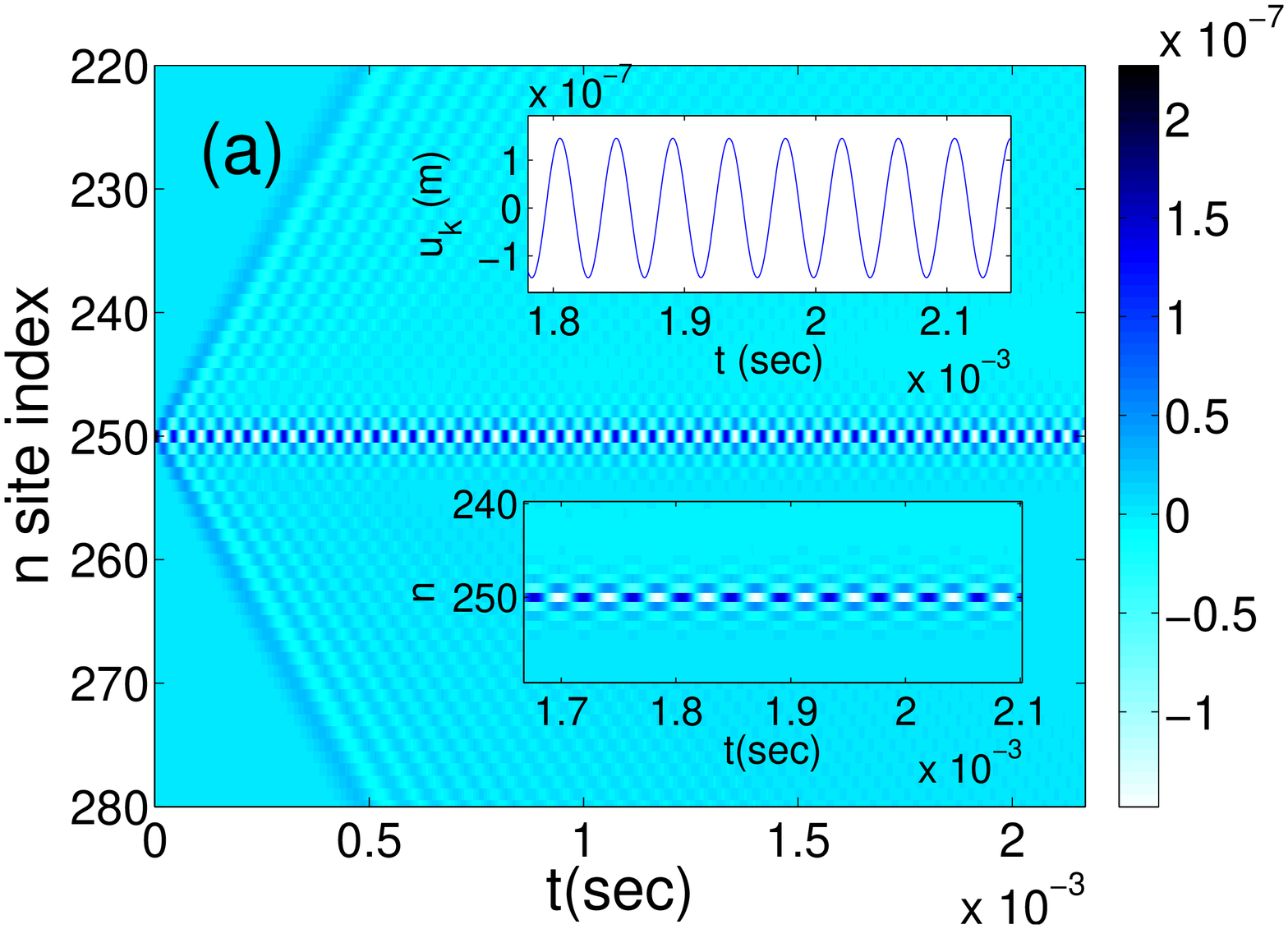}
\includegraphics[width=8cm]{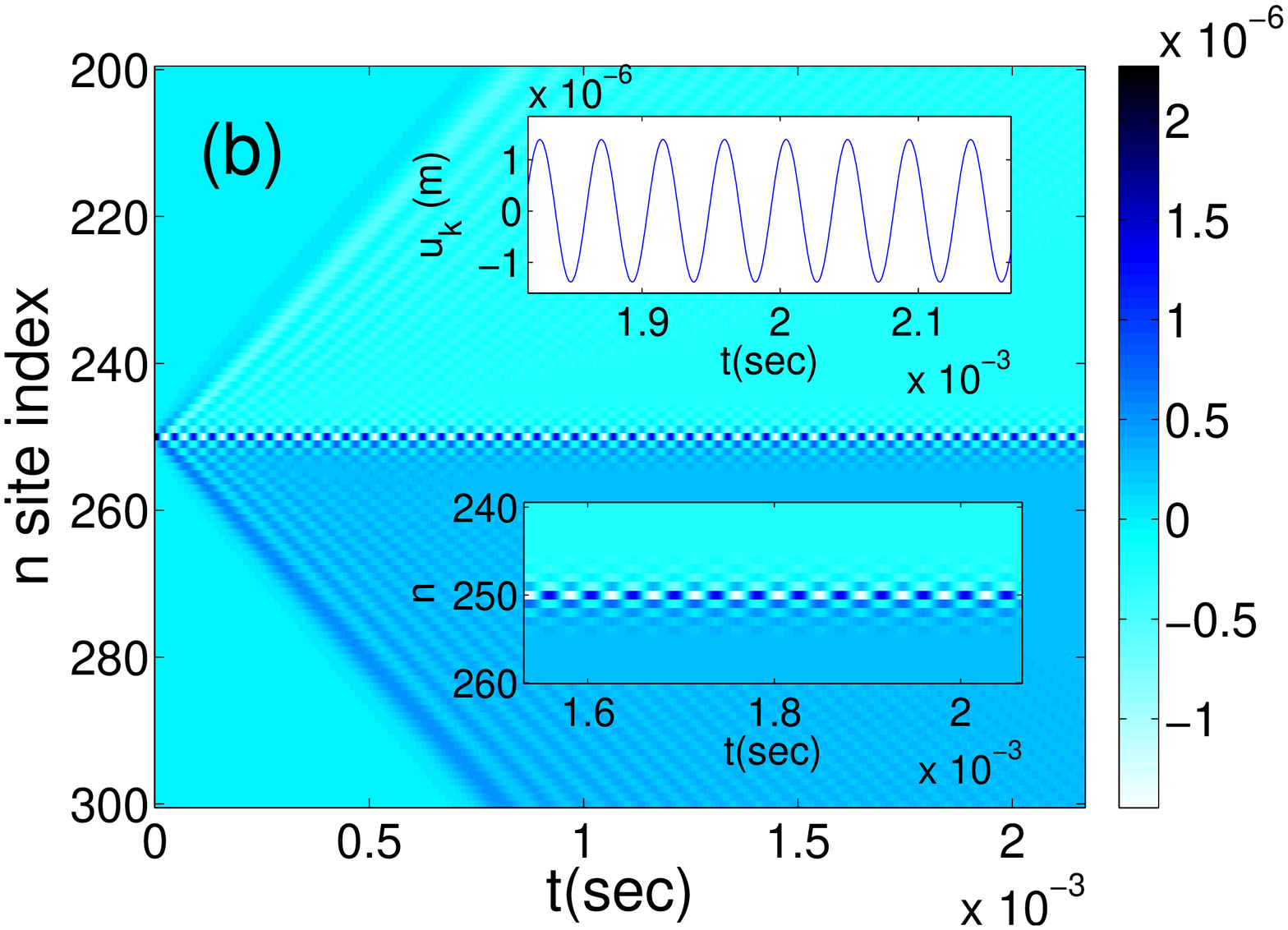}
\caption{(Color online) Spatiotemporal evolution of the displacements of the beads with initial conditions  (a) $u_{k}(0)=\delta_{1}/10$ and $u_j = 0$ for all $j \neq k$ and (b) $u_{k}(0)=\delta_{1}$ and $u_j = 0$.  Top insets: The displacements of the impurity bead (which lies at the $k$th site). Bottom inset: magnifications of the spatiotemporal evolution near the impurity site.  Observe that after an initial (transient) stage of energy shedding in the form of sound waves, a nonlinear localized breathing mode forms in the neighborhood of the defect.  
}
\label{generation}
\end{figure}

We generated long-lived localized modes using two proof-of-principle simulations.  In the first, we considered a relatively weak initial displacement of the impurity site, $u_{k}(0)=\delta_{1}/10$, which we show in Fig.~\ref{generation}(a).  In the second, we considered a relatively strong initial displacement of the impurity site, $u_{k}(0)=\delta_{1}$, which we show in Fig.~\ref{generation}(b).  In both cases, we used the precompression and material parameters discussed above. 
Observe that for the strong initial displacement, the nonlinearity causes a substantial distortion of the chain. The frequency of the oscillations of the impurity site was about $23.25$ kHz and $\max(u_{k}/\delta_{1}) \approx 0.065$ for the first simulation.  For the second simulation, we observed a frequency of about $22.47$ kHz and $\max(u_{k}/\delta_{1}) \approx 0.5$.  Both sets of results are in excellent agreement with the continuation analysis discussed above, as that gave frequencies of $23.26$ kHz and $22.41$ kHz for these particular values of $\max(u_{k}/\delta_{1})$.  These simulations therefore clearly illustrate the excitation of the previously analyzed NLMs.

\section{Monoatomic Granular Chain with Two Impurities}

\subsection {Harmonic Potential Approximation (Linear Analysis)}

\begin{figure}[tbp]
%\centering 
\includegraphics[width=8cm]{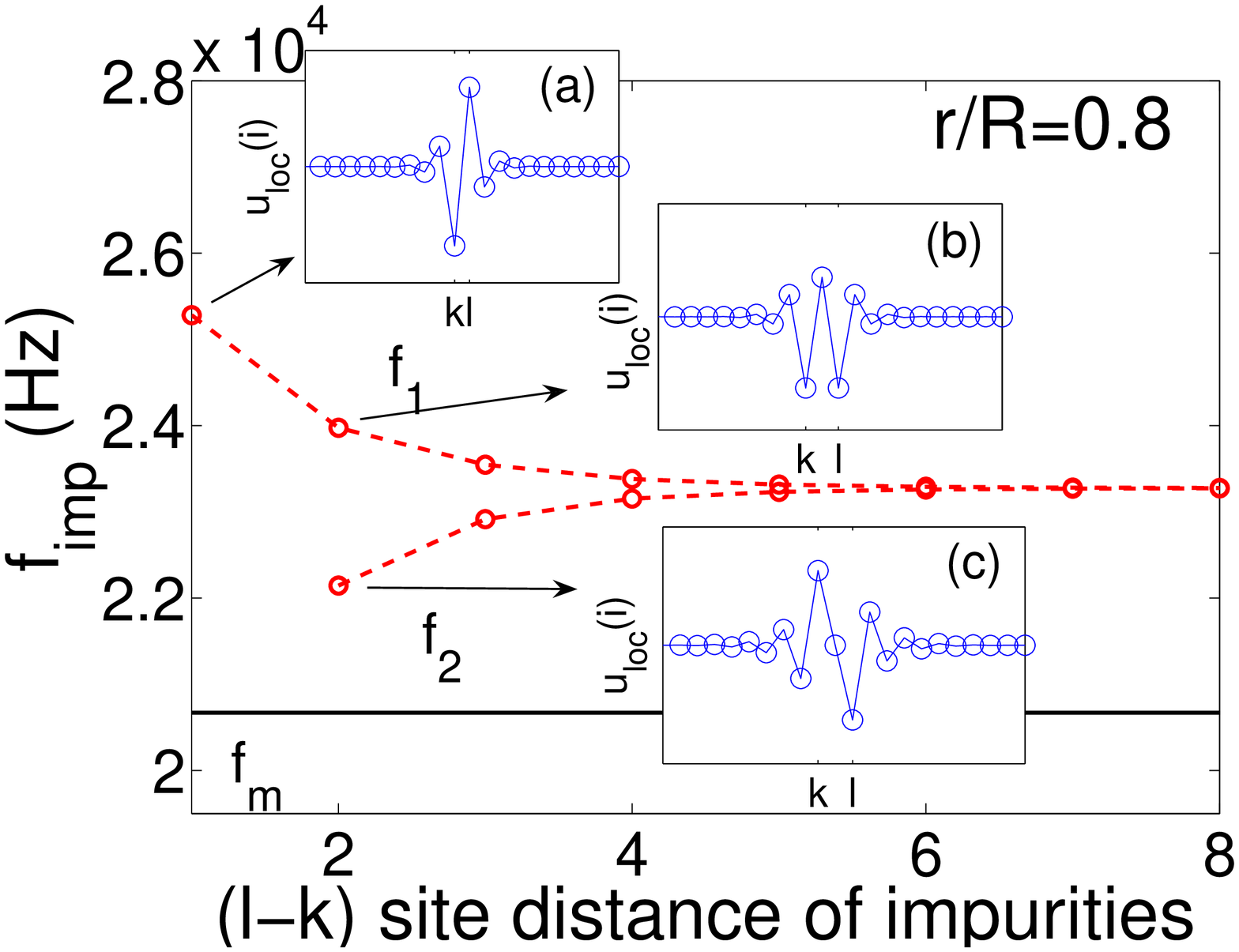}
\includegraphics[width=8cm]{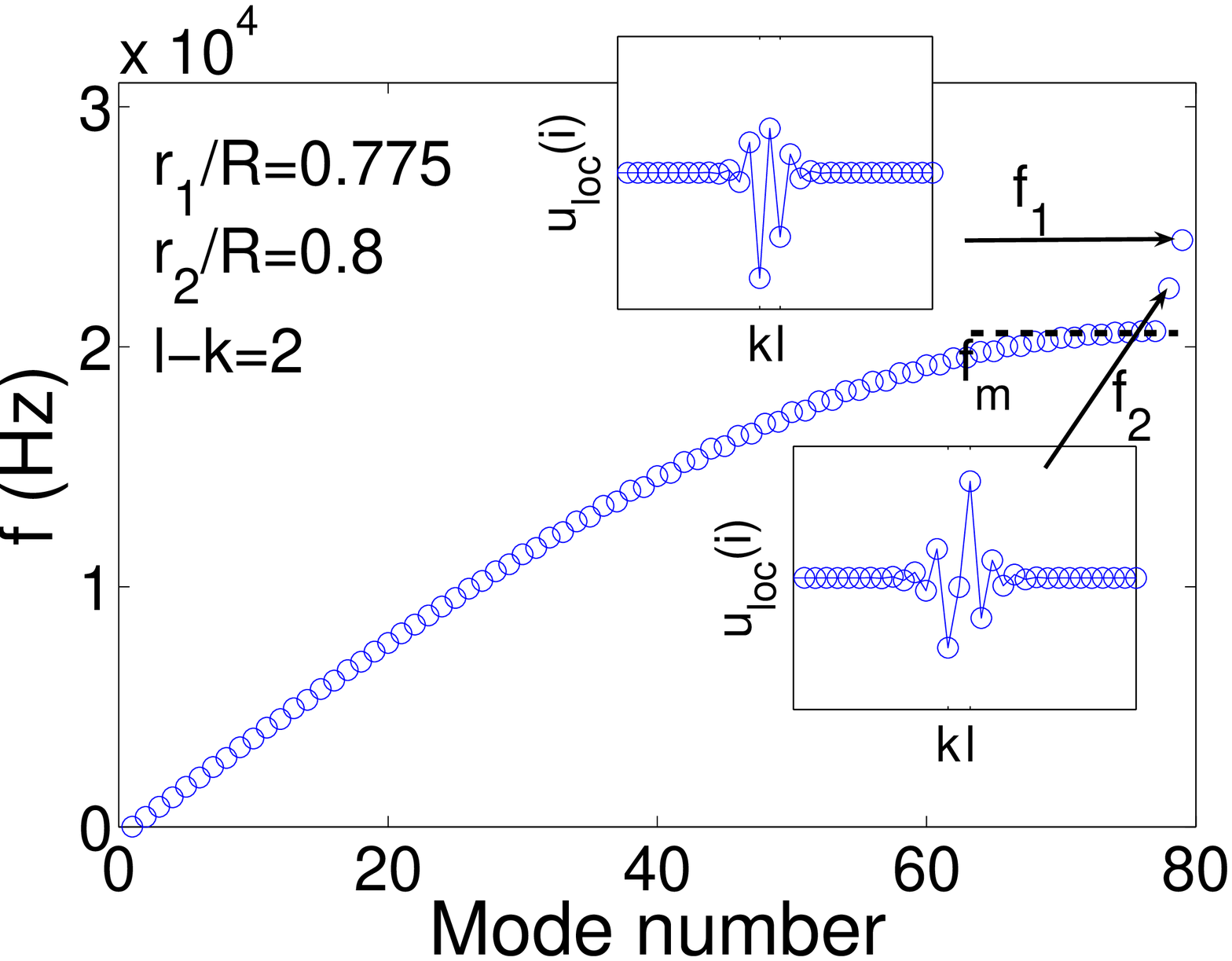}
\caption{(Color online) Left Panel: The frequencies of the localized modes generated by two identical impurities as a function of the distance (number of sites) between them.  The insets show corresponding 
localized modes for (a) $l - k = 1$, for which the impurities are in contact and (b,c) $l - k = 2$.  As the impurities are placed increasingly far apart, we see that $f_1 - f_2\rightarrow 0$. Right Panel: The normal mode frequencies of a granular crystal with two impurities, with $l - k = 2$, of radii $r_1 = 0.775R$ and $r_2 = 0.8R$.  
}
\label{2def_site_distance}
\end{figure}

We now consider a crystal with two impurities, located at sites $k$ and $l$.  As before, we begin with an analysis of the linear spectrum.  
We first examine the case of two identical impurities, for which we consider impurity beads made of the same material (stainless steel) as the ones of the host chain (which is again a perfect crystal) but of smaller radius ($r_1 = r_2 = r = 0.8 R$).  Figure \ref{2def_site_distance} shows the frequencies that correspond to localized modes, which we obtained using the harmonic approximation, as a function of the distance between the impurities. Interestingly, for this value of radius ratio, when the impurity beads are in contact ($l - k = 1$), the phonon spectrum has just a single localized mode with frequency $f_{imp} \approx 25.28$ kHz. As shown in inset (a) in the left panel of Fig.~\ref{2def_site_distance}, the corresponding mode is antisymmetric. 
Studying the phonon spectrum of the case $l - k = 1$, as a function of the radius ratio $r/R$, we found that for $r/R<0.8$ a symmetric mode
leaves the phonon band and becomes progressively more localized as the radius ratio is decreased. In particular, for $r/R=0.4$ the phonon spectrum
contains two localized modes, an antisymmetric with $f_1\approx 62.26 kHz$ and a symmetric with $f_2\approx 38.48 kHz$.
When $l - k \geq2$, there are two localized modes even for $r/R=0.8$.
In particular, for $l - k = 2$, the frequency of the symmetric mode [inset (b) in the left panel of Fig.~\ref{2def_site_distance}] is $f_1 \approx 23.98$ kHz, and the frequency of the antisymmetric mode  [inset (c) in the left panel of Fig.~\ref{2def_site_distance}] is $f_2 \approx 22.14$ kHz.  As the two impurities are placed farther apart, $f_1 - f_2 \rightarrow 0$, and $f_1\,, f_2 \rightarrow f_{imp} \approx 23.28$ kHz, which is the frequency of a single impurity localized mode (see the discussion in the previous section).

We now consider the phonon spectrum for the case of two different impurities with separation $l - k = 2$ and bead radii $r_1 = 0.775R$ and $r_2 = 0.8R$.  (As before, the impurity beads are made of the same material as those in the host chain.)  As one can see by comparing the insets in the left and right panels of Fig.~\ref{2def_site_distance}, even a slight difference in the radii of impurity beads results in an asymmetric modification of the corresponding localized modes.  In this case, we also obtain slightly larger mode frequencies: $f_1 \approx 24.45$ kHz and $f_2 \approx 22.44$ kHz.

\subsection{Continuation And Stability Analysis For Two Identical Impurities}

\begin{figure}[tbp]
%\centering 
%\includegraphics[width=8cm]{bifurcation2d_incontact_r_8_10.eps}
%\includegraphics[width=8cm]{maxeigenvalue2def_8_10.eps}
\includegraphics[width=8cm]{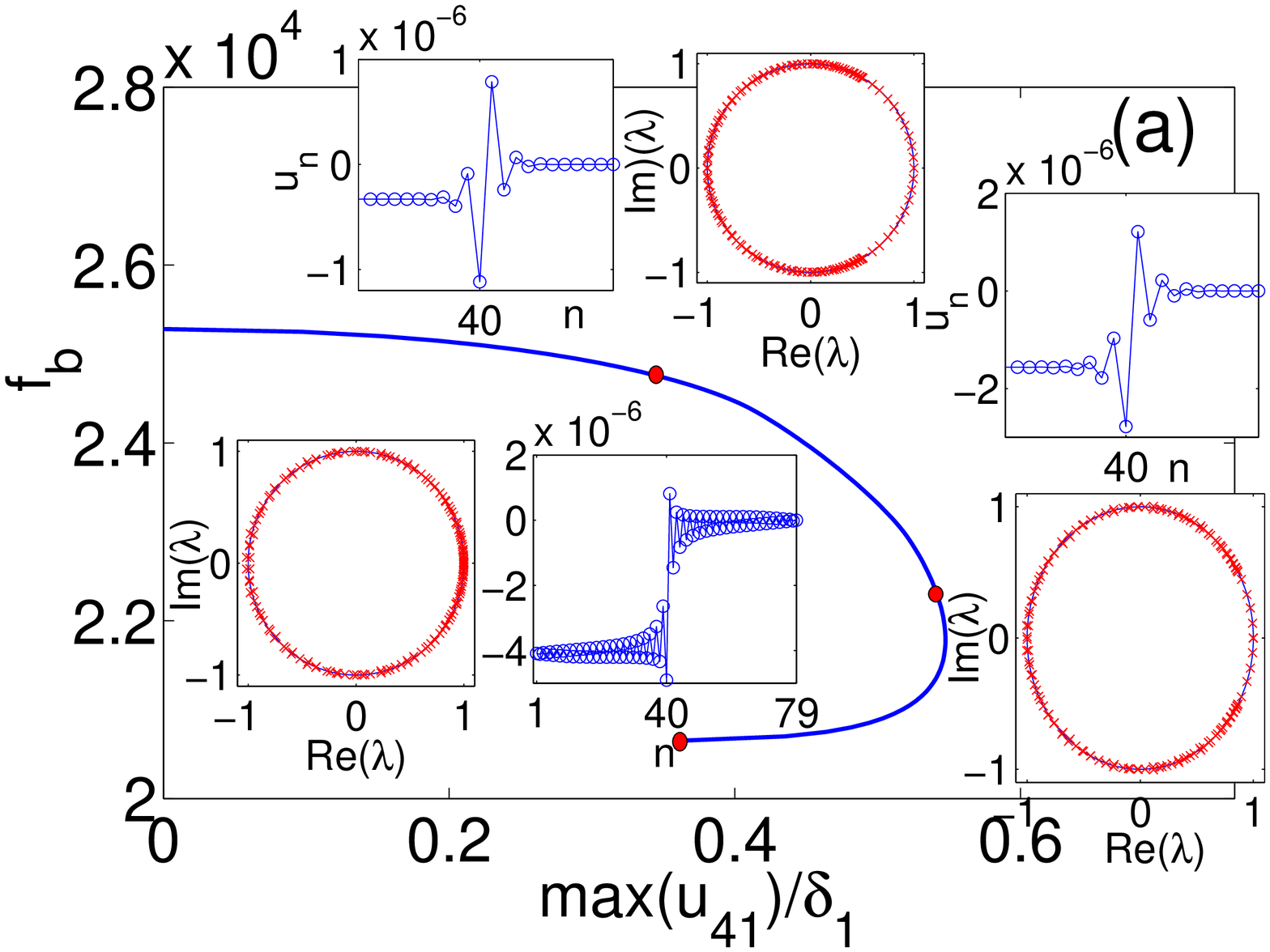}
\includegraphics[width=8cm]{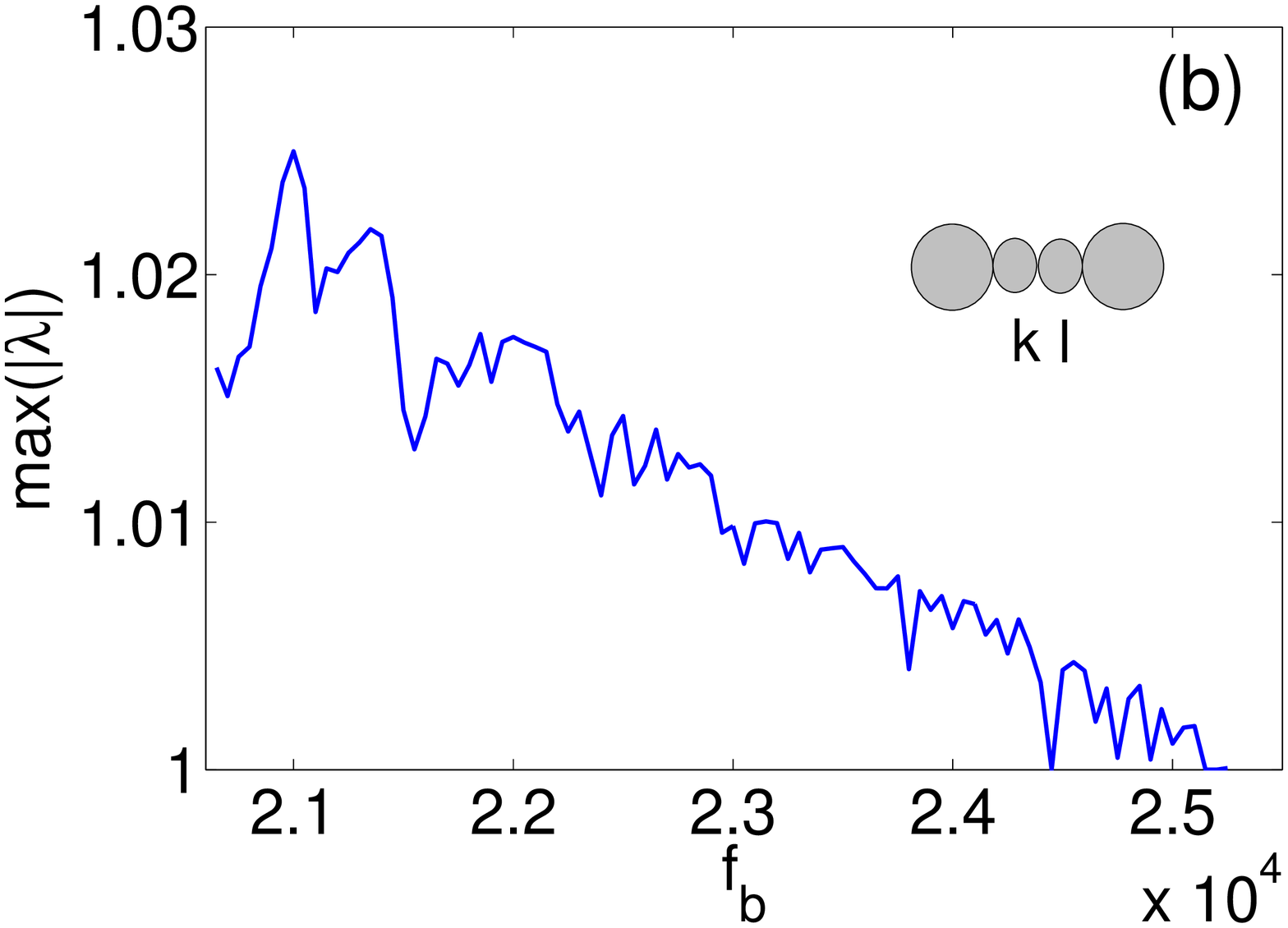}
\caption{(Color online) Left panel: Continuation diagram of the nonlinear impurity modes for the case of two identical impurity beads in contact ($l - k = 1$).  One set of insets shows the profiles of the localized modes with frequencies $f_1 =24.75$ kHz, $f_2=22.55$ kHz, and $f_m=20.65$ kHz, and the other set shows the corresponding locations of their Floquet multipliers $\lambda$ in the complex plane.  Right panel: The maximum of the absolute value of Floquet multipliers as a function of the frequency $f_{b}$ of the nonlinear impurity mode. 
}
\label{contin1}
\end{figure}

We show the results of our parameter continuation for the case of two identical impurity beads in contact ($l - k = 1$), with radius ration $r/R=0.8$ in Fig.~\ref{contin1}.  We find essentially the same phenomenology as we obtained for granular crystals with a single impurity.  That is, we obtain a family of weakly-oscillatory unstable localized solutions due to finite size effects (the magnitudes of the deviations of the unstable eigenvalues from the unit circle are smaller than $0.025$).  Again as before, the modes become wider and the characteristic dc distortion of the chain becomes larger as one approaches the frequency $f_m$. We show three examples of this family of localized solutions (with frequencies $f_1=24.75$ kHz, $f_2=22.55$ kHz, and $f_m=20.65$ kHz) in insets of Fig.~\ref{contin1}(a).  In the rest of the insets, we show the locations of 
their corresponding Floquet multipliers in the complex plane.

\begin{figure}[tbp]
%\centering 
%\includegraphics[width=8cm]{2d_alter_stable.eps}
%\includegraphics[width=8cm]{2d_alter_unstable_new.eps}
%\includegraphics[width=8cm]{2d_alter_stable_eig.eps}
%\includegraphics[width=8cm]{2d_alter_unstable_eig.eps}
\includegraphics[width=8cm]{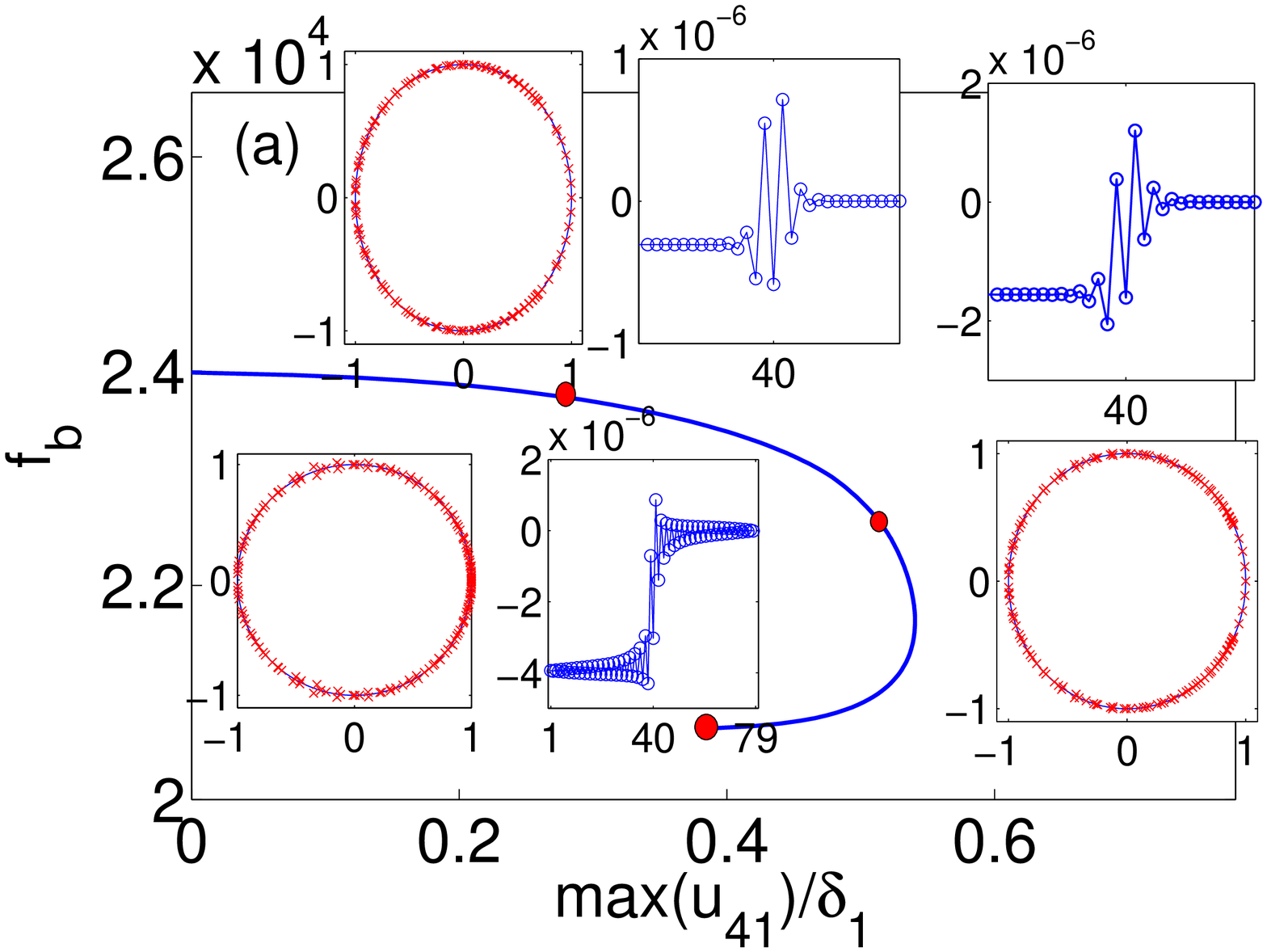}
\includegraphics[width=8cm]{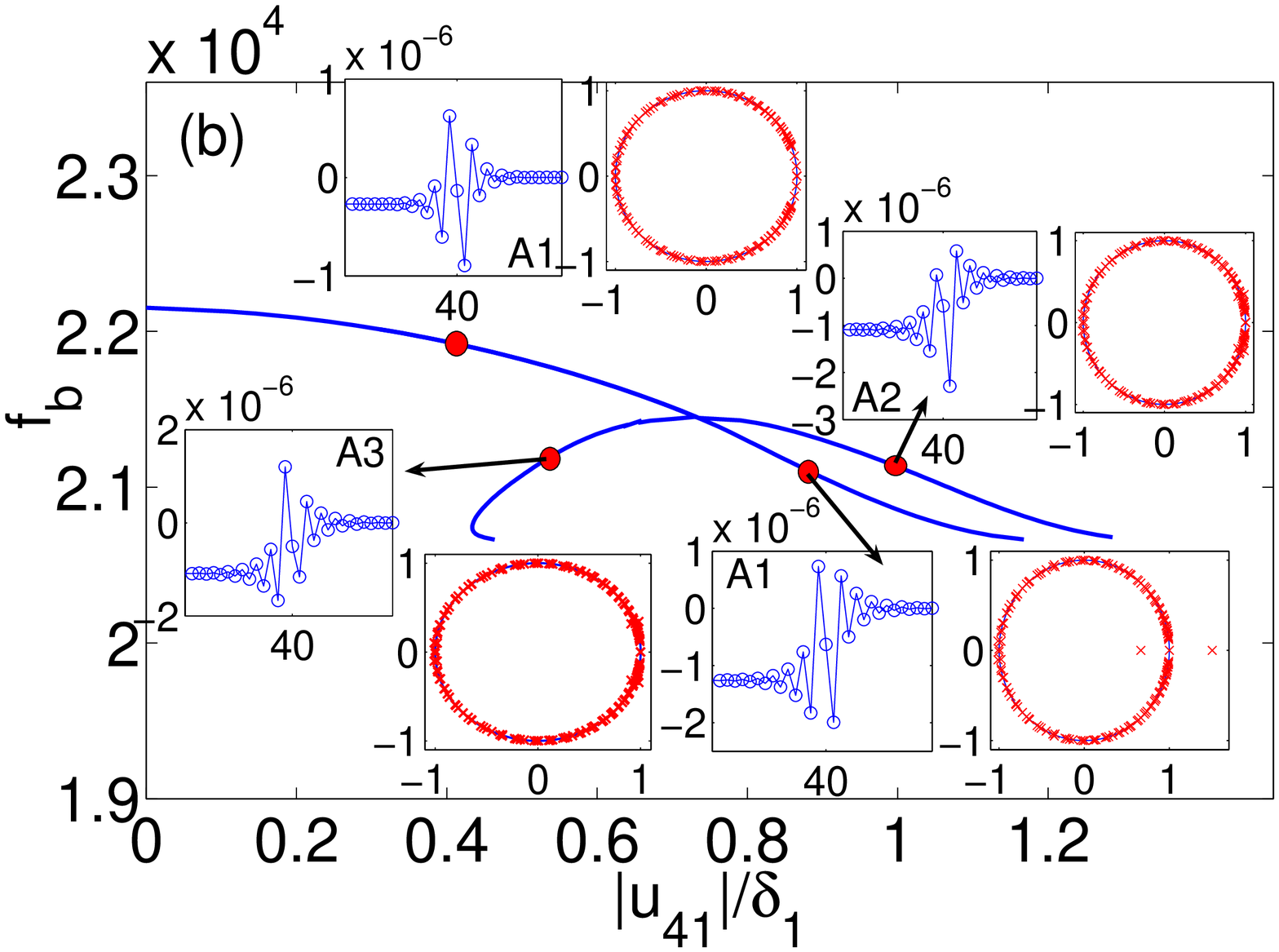}
\includegraphics[width=8cm]{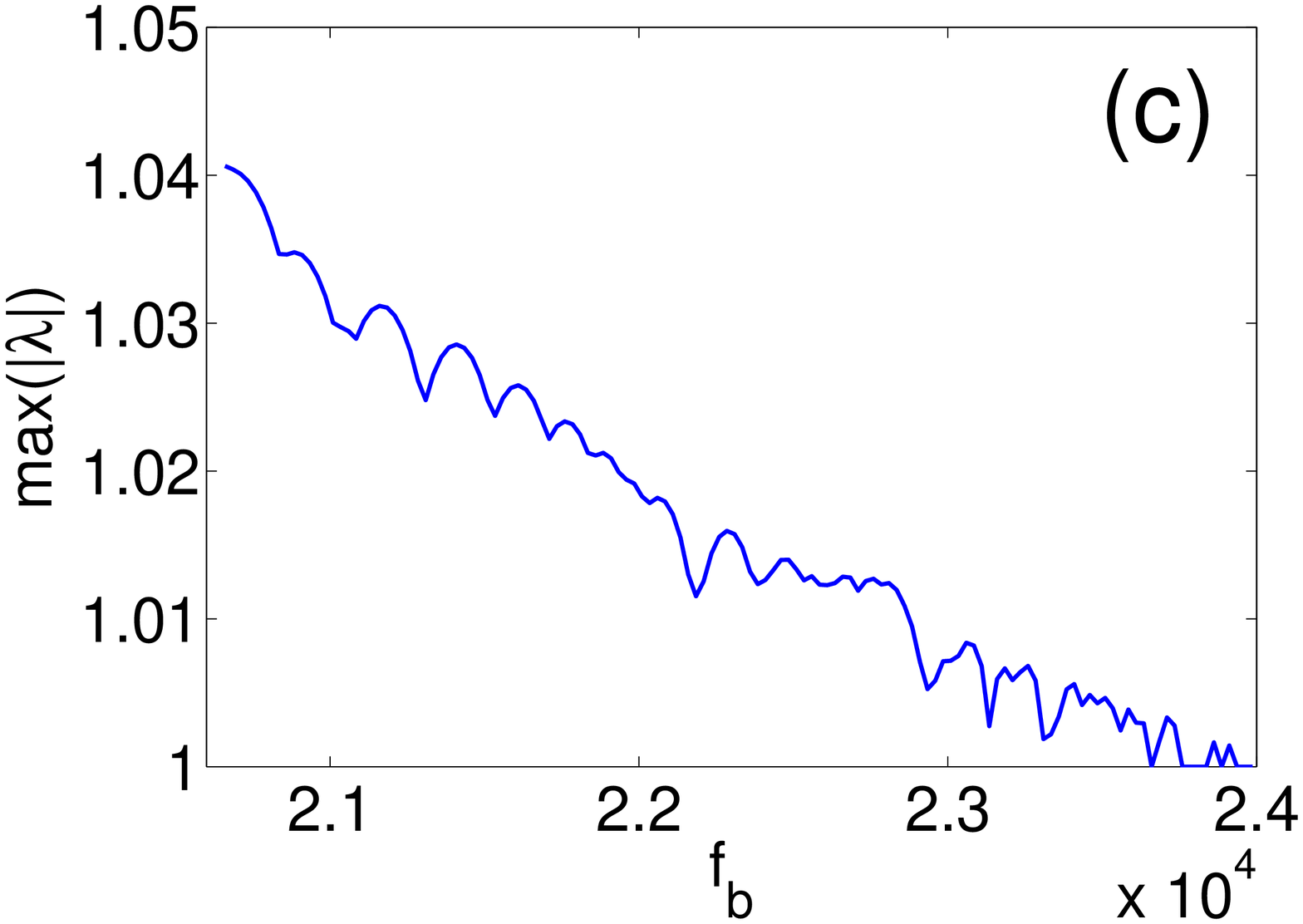}
\includegraphics[width=8cm]{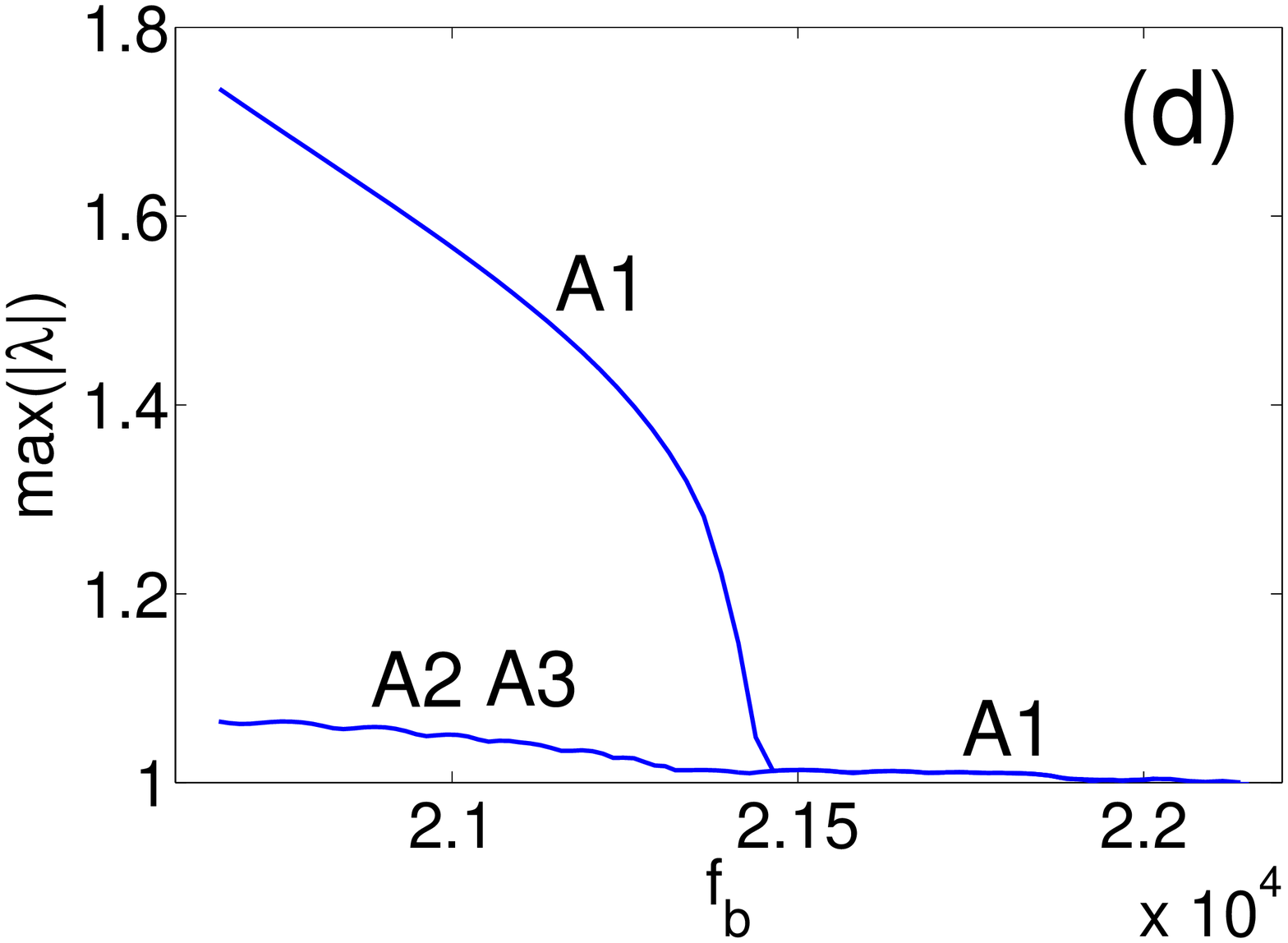}
\caption{(Color online)(a) Continuation diagram for the nonlinear impurity modes that originate from the symmetric linear mode of a granular crystal with two impurities that are separated by one bead (i.e., $l - k = 2$).  One set of insets shows the profiles for the localized modes with frequencies $f=23.68$ kHz, $f=22.26$ kHz, and $f=20.66$ kHz.  The other set shows the locations in the complex plane of their corresponding Floquet multipliers $\lambda$. (b)
Continuation diagram for the antisymmetric solution branch ($A_1$), showing the generation of asymmetric branches ($A_2$ and $A_3$) that originate from the bifurcation at $f \approx 21.44$ kHz.  The insets show the wave profiles (and their Floquet multipliers) of the antisymmetric localized mode before the bifurcation ($f_b=21.94$ kHz), the antisymmetric mode after the bifurcation ($f_b=21.11$ kHz), and the asymmetric mode with frequency $f_b=21.11$ kHz.
(c) The maximum of the absolute value of Floquet multipliers as a function of the frequency of the nonlinear impurity mode $f_{b}$ for the symmetric branch.  (d)
Same as (c), but for antisymmetric ($A_1$) and asymmetric ($A_2$, $A_3$) branches. 
}
\label{contin2}
\end{figure}

Now consider a granular crystal with two impurities that are not in contact.  More specifically, we focus on the prototypical case of $l - k = 2$ and $r/R=0.8$.  As indicated above, the corresponding phonon spectrum contains two localized modes: a symmetric one, shown in inset (b) of the left panel of Fig.~\ref{2def_site_distance}, and an antisymmetric one, shown in inset (c) of the left panel of Fig.~\ref{2def_site_distance}.  In Fig.~\ref{contin2}, we show continuation diagrams, which display
the frequencies of NLMs as a function of the maximum displacement of one 
of the impurity beads ($l = 41$, in a chain with $N = 79$ beads) normalized to the characteristic (precompression-induced) displacement $\delta_{1}$.  First, we examine the family of solutions that arises from the symmetric linear mode at $f_1 \approx 23.98$ kHz, see Fig.~\ref{contin2} (a).  
Stability analysis demonstrates the presence of a weak oscillatory instability as in the case of a single and two in-contact impurities.
Now consider the nonlinear localized solutions that bifurcate from the antisymmetric linear mode, for which $f_2 \approx 22.14$ kHz.  This family of solutions, which corresponds to the branch $A_1$ of the continuation diagram in panel (b) of Fig.~\ref{contin2}, is initially weakly unstable (due to the finite-size effects discussed previously).
At $f\approx21.56kHz$, a pair of Floquet multipliers leave the phonon bands. The corresponding eigenmodes are symmetric and become progressively localized \cite{Baesens} as the frequency decreases. At $f\approx21.44kHz$, these two localized modes, collide at the $(+1,0)$ point of the unit circle, giving rise to a strong instability (called harmonic instability) which is connected to a bifurcation of the corresponding NLM.
Two new families of solutions (branches $A_2$ and $A_3$) emerge from this bifurcation which, excluding the kink-shaped distortion of the system, 
are symmetric to each other and weakly unstable due to finite-size effects. 
Thus, this bifurcation is somewhat reminiscent of a pitchfork 
bifurcation \cite{Iooss}. In the case of the newly formed branches
$A_2$ and $A_3$
past the bifurcation point, and particularly at $f\approx21.34kHz$, the formed localized pair of eigenmodes enters the band of eigenvalues associated with extended perturbations giving rise to a new oscillatory instability.
This kind of instability although size-dependent, in contrast to the 
oscillatory instability caused by the collision
of two extended modes, 
persists even in the limit of an infinite system \cite{Johannson}.
 
It is worth noting that the setting of granular chains with two next-nearest-neighbor impurities (i.e., with $l - k = 2$) is reminiscent of double-well configurations in other contexts.  For example, both defocusing and focusing nonlinear Schr{\"o}dinger (NLS) equations with double-well potentials are known to exhibit ``symmetry breaking'' bifurcations like the one discussed above \cite{wei12us}.  (The defocusing case is relevant to the present setting.)  Moreover, these bifurcations have even been observed experimentally  in both optical \cite{weiexp} and atomic systems \cite{markus}.

\begin{figure}[tbp]
%\centering 
%\includegraphics[width=8cm]{dynamic_unstable_antisymmetric.eps}
%\includegraphics[width=8cm]{dynamic_unstable_asymmetric.eps}
\includegraphics[width=8cm]{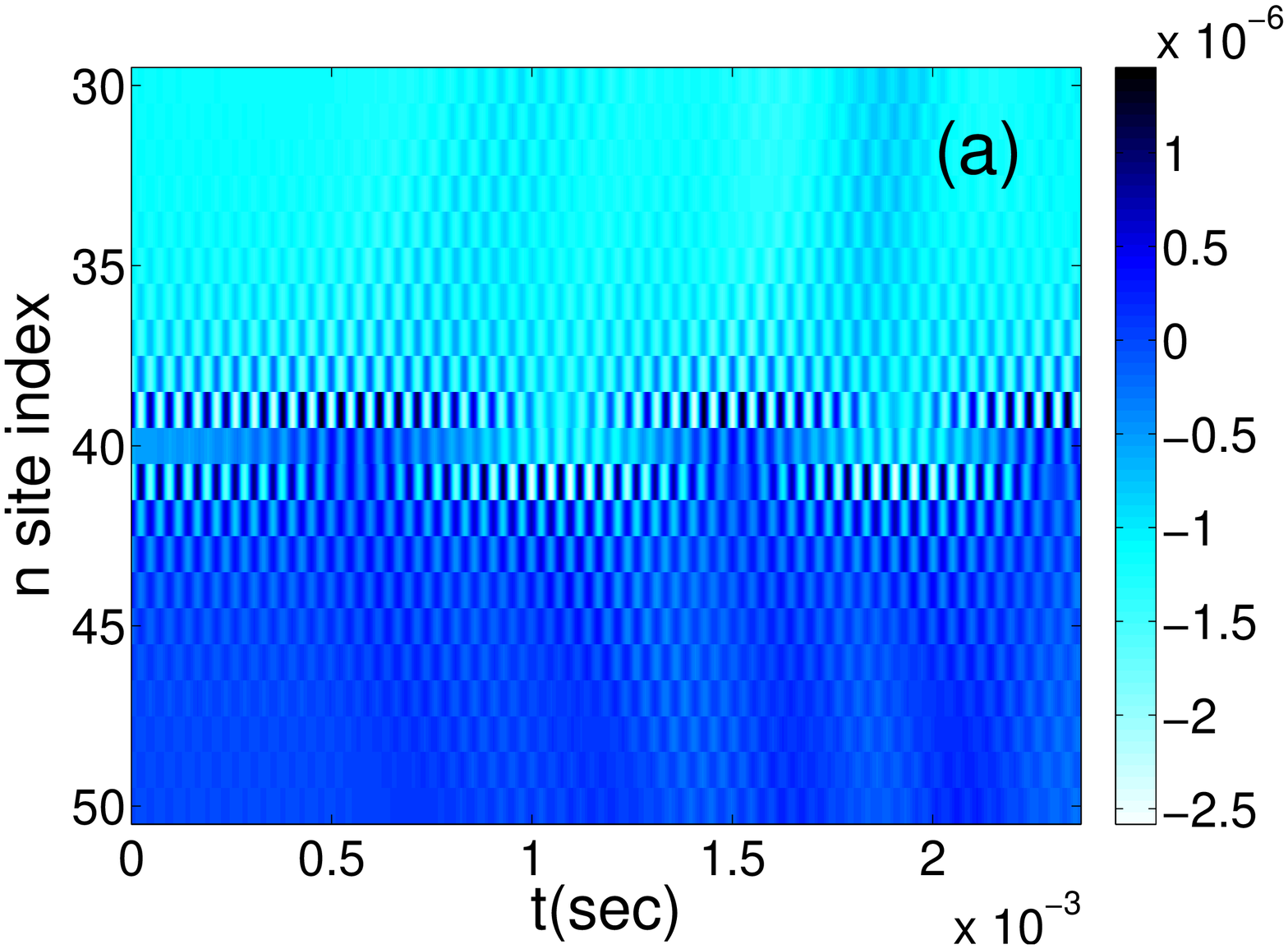}
\includegraphics[width=8cm]{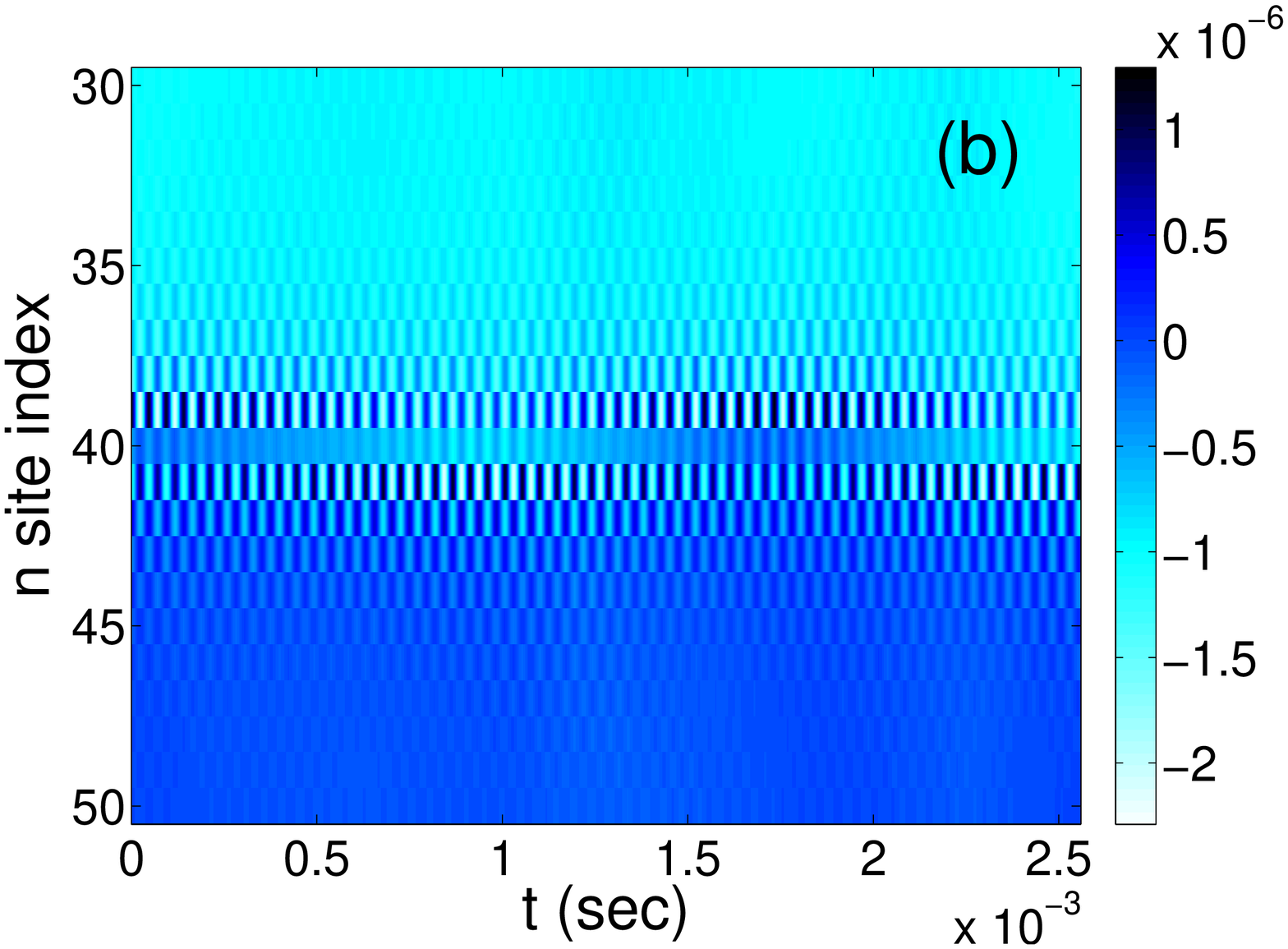}
\caption{(Color online) Spatiotemporal evolution of the bead displacements for (a) the harmonically unstable antisymmetric localized mode with $f_b=21.214$ kHz (for which ${\rm max} |\lambda| \approx 1.437$) and (b) for the weakly oscillatory unstable asymmetric localized mode with the same frequency (for which ${\rm max} | \lambda | \approx 1.032$).}
\label{dyn_instab}
\end{figure}

Examining the temporal dynamics of the unstable antisymmetric mode evinces the symmetry-breaking phenomenon.  To trigger the relevant instability, we use the wave given by the sum of the unstable solution with $f \approx 21.214$ kHz and the corresponding unstable localized eigenfunction as an initial condition in the full nonlinear equations of motion.  
Its dynamical evolution, which we show in Fig.~\ref{dyn_instab}(a), reveals the ``symmetry breaking'' at $t \approx 0.4$ ms.  This is followed by alternating oscillations between the two impurity sites 
($A_2$ and $A_3$ asymmetric modes). 
As illustrated in Fig.~\ref{dyn_instab}(b), the dynamic evolution of the 
weak oscillatory instability of the asymmetric modes is 
somewhat similar.  As pointed out above, such dynamics is 
reminiscent of theoretical \cite{wei12us} and experimental \cite{weiexp} 
observations of the instability manifestation in NLS equations with double
well potentials.

\subsection{Continuation And Stability Analysis For Two Distinct 
Impurities}

As we discussed above, we observe a pitchfork-like bifurcation, indicating the emergence of two families of asymmetric localized solutions 
if the two impurities in the granular chain are identical.  Employing our analogy with NLS equations, this means that the observed dynamics corresponds to that obtained in a symmetric ``double well'' that can be considered as being 
induced by the identical, next-nearest-neighbor impurities. 
A natural generalization is to consider how the phenomenology we studied above changes when these impurities are distinct.  In this case, the ``double well'' becomes asymmetric, and (from bifurcation theory) one expects to see something analogous to what is sometimes called an imperfect pitchfork bifurcation \cite{Iooss}.  Such a scenario, which has been observed in NLS equations with asymmetric double wells \cite{wei12us1}, involves an asymmetric perturbation of the pitchfork structure, resulting in a saddle-node bifurcation and an isolated branch of solutions.

\begin{figure}[tbp]
%\centering 
\includegraphics[width=8cm]{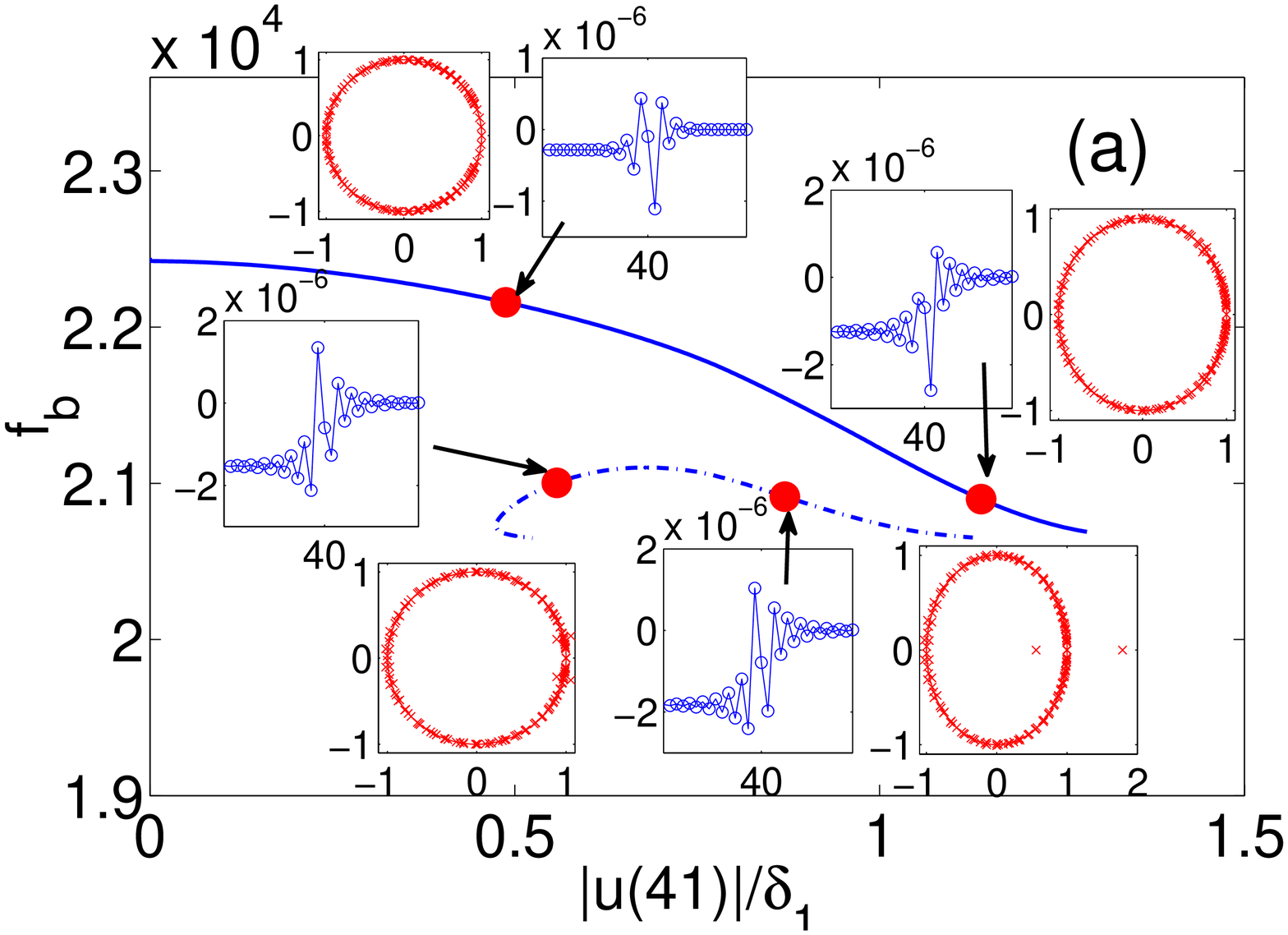}
\includegraphics[width=8cm]{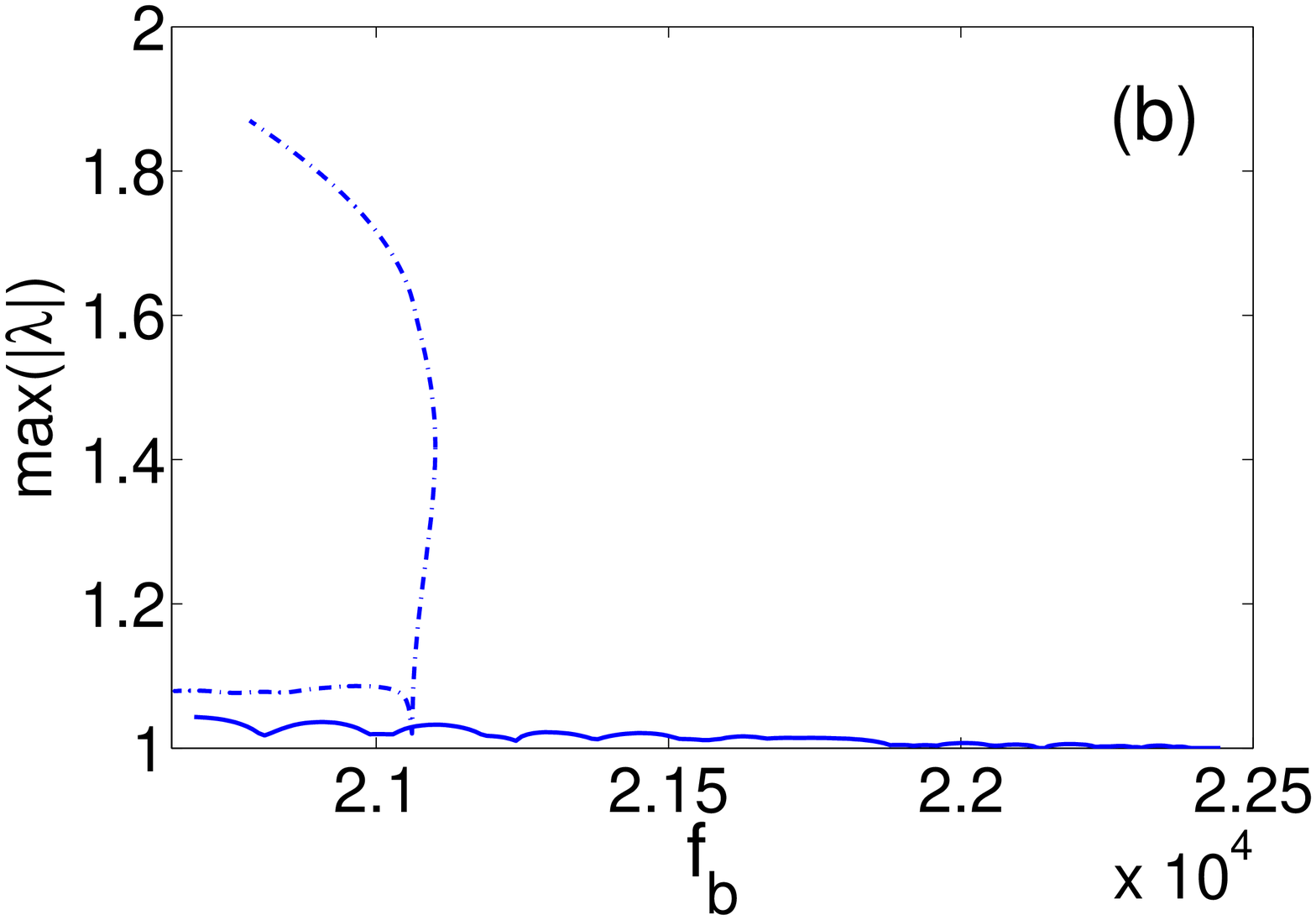}
\caption{(Color online) (a) Continuation diagram for the nonlinear impurity mode that originates from $f_2 \approx 22.44$ kHz.  This diagram also shows two additional families of solutions that collide at $f \approx 21.1$ kHz.  The insets show the spatial profiles (and associated Floquet multipliers $\lambda$) of the localized modes at $f_b=22.16$ kHz and $f_b=20.9$ kHz.  In the latter case, we show an example from each branch. 
(b) The maximum of the absolute value of Floquet multipliers as a function of the frequency of the nonlinear impurity mode $f_{b}$ for the all of the branches. Observe that the isolated branch that stems from the linear impurity mode is only weakly unstable in the frequency interval that we probed, whereas the two branches that arise via the saddle-node bifurcation and exist for $f < 21.1$ kHz encompass a strongly unstable family of solutions and a weakly unstable one.}
\label{saddlenode}
\end{figure}

To observe this breakdown and obtain the associated modified bifurcation picture, we consider the 
case of two distinct impurities with slightly different radii (namely, $r_1 = 0.775R$ and $r_2  =  0.8R$) on the prototypical next-nearest-neighbor case of $l - k = 2$.  Recall that we showed the normal mode frequencies for this configuration in the right panel of Fig.~\ref{2def_site_distance}.  We show the continuation and the stability diagrams for the family of solutions originating from $f_2 \approx 22.44$ kHz in Fig.~\ref{saddlenode}.  (We omit the continuation of the solutions that emerge from the 
linear mode with frequency $f_1$ because it resembles the one shown in the left panel of Fig.~\ref{contin2}.)  As suggested above, this amounts to a
saddle-node bifurcation. 
The branches of solutions that are analogous to the $A_1$ and $A_3$ branches from Fig.~\ref{contin2} collide at a critical value $f \approx 21.1$ kHz and disappear. 
Linear stability analysis shows that one colliding branch consists of strongly harmonically unstable solutions and the other consists of weakly oscillatory unstable solutions. The isolated branch, which occurs here because we have broken the perfect pitchfork and arises from the frequency $f_2$ of the linear limit, is only weakly unstable instead of exhibiting the strong instability we showed in Fig.~\ref{contin2}. Once again, this is reminiscent of the NLS phenomenology observed in \cite{wei12us1}.

\section{Conclusions and Future Challenges}

In conclusion, we have 
investigated the formation of localized modes due to the interplay of nonlinearity and disorder (i.e., the presence of defects) in granular crystals.  While previous research has been dedicated to the computational and experimental identification of such modes \cite{sen08,Job}, we believe that the present work is the first one to identify such modes in a {\it numerically exact} form and 
offer a systematic analysis of their linear stability.  We have argued that these localized modes are likely to
be absent in monoatomic chains, and we have illustrated that the inclusion of even a single defect can produce a linear defect mode whose interplay with nonlinearity generates a full branch of such solutions.  We have demonstrated that these waves are robust, and it should be straightforward to generate them in experiments for a wide range of initial conditions.  We extended our single-defect investigation to multi-defect settings, for which we examined the prototypical situation of crystals with two defect sites.  We identified a much richer phenomenology in such situations, as we observed families of both near-symmetric and near-antisymmetric states.  We subsequently demonstrated that the latter yields additional families of asymmetric solutions through pitchfork-like bifurcations, which we analyzed for defect pairs containing both identical and distinct impurities.

Although this paper presents some of the first steps towards a systematic
understanding and classification of localized breathing modes
in granular crystals, there are numerous interesting future directions that should be pursued.  First, one can consider progressively larger numbers of defects.  By analogy with NLS settings \cite{todd}, we expect this to reveal not only a wealth of additional phenomenology (e.g., new families of waves and more complicated bifurcation structures) but also implications for how things progress toward the
infinite defect limit.  If all defects occur only as next-nearest-neighbors, we would obtain a perfect diatomic (two species) crystal in this limit.  Such a crystal is naturally expected to support intrinsic localized mode solutions in the band gap between its acoustic and optical bands.  Indeed, our preliminary results indicate that such gap breather solutions do indeed arise.  The detailed examination of such modes, and their higher-dimensional generalizations (including possibly gap vortex modes \cite{kivsharus}), are among our future goals.

\section*{Acknowledgements}

PGK gratefully acknowledges support from NSF through NSF-DMS-0349023
(CAREER) and NSF-DMS-0806762, as well as from the Alexander von
Humboldt Foundation. The work of IGM and MK was partially supported by NSF and ACS(PRF). CD thanks NSF CAREER and NSF CMMI for support of this project.

\end{document}